\documentstyle[aps,twocolumn,floats,prd,psfig]{revtex}

\newcommand{\ApJ}{Astrophys. J.}

\newcommand{\PRD}{Phys. Rev. D}
\newcommand{\MNRAS}{MNRAS}

\newcommand{\aut}[2]{{#2.\ #1}}
\newcommand{\refs}[6]{#2, {\bf #3} {#4} (#5)}

\newcommand{\amp}{and }

\newcommand{\da}{d_A}

\newcommand{\tot}{{\rm t}}
\newcommand{\ng}{{\rm NG}}
\newcommand{\hal}{{\rm h}}

\newcommand{\lin}{{\rm lin}}
\newcommand{\halo}{{\rm h}}
\def\sun{\hbox{$\odot$}}

\newcommand{\bfl}{{\mathbf{l}}}
\newcommand{\dirac}{{\rm D}}
\newcommand{\bp}{{\cal C}}
\newcommand{\shell}{{\rm s}}

\newcommand{\bfx}{{\mathbf{x}}}

\newcommand{\veck}{{\bf k}}
\newcommand{\vecl}{{\bf l}}
\newcommand{\vecr}{{\bf r}}
\newcommand{\vecx}{{\bf x}}

\newcommand{\rms}{{\it rms}}
\newcommand{\cmb}{\Theta}
\newcommand{\s}{{\rm s}}
\newcommand{\vecla}{{{\bf l}_1}}
\newcommand{\veclb}{{{\bf l}_2}}
\newcommand{\veclc}{{{\bf l}_3}}
\newcommand{\vecld}{{{\bf l}_4}}
\newcommand{\intl}[1]{\int {d^2 {\bf l}_#1 \over (2\pi)^2}}

\newcommand{\Cov}{{\rm Cov}}

\newlength{\tskip}
\newlength{\colwidth}

\newcommand{\beq}{\begin{equation}}
\newcommand{\eeq}{\end{equation}}
\newcommand{\beqa}{\begin{eqnarray}}
\newcommand{\eeqa}{\end{eqnarray}}
\newcommand{\bi}{B_{l_1 l_2 l_3}}

\newcommand{\deld}{\delta^{\rm D}}
\newcommand{\bn}{\hat{\bf n}}

\newcommand{\bk}{\hat{\bf k}}
\newcommand{\rad}{r}    % comoving radial distance

\newcommand{\dop}{{\rm dop}}
\newcommand{\sky}{{\rm sky}}

\newcommand{\se}{{\rm S}}
\newcommand{\ri}{{\rm ri}}
\newcommand{\len}{\phi}

\newcommand{\Ylmn}{Y_{l}^{m}}
\newcommand{\alm}[1]{a_{l_#1 m_#1}}

\newcommand{\dsz}{{\rm kSZ}}
\newcommand{\sz}{{\rm SZ}}
\newcommand{\nm}{\frac{d\bar{n}}{dM}}
\newcommand{\nma}{\frac{d\bar{n}}{dM_1}}
\newcommand{\nmb}{\frac{d\bar{n}}{dM_2}}
\newcommand{\nmc}{\frac{d\bar{n}}{dM_3}}
\newcommand{\nmd}{\frac{d\bar{n}}{dM_4}}
\newcommand{\sn}{\frac{{\rm S}}{{\rm N}}}
\newcommand{\noise}{{\rm noise}}
\newcommand{\g}{{\rm G}}

\begin{document}
\twocolumn[\hsize\textwidth\columnwidth\hsize\csname
@twocolumnfalse\endcsname

\title{Non-Gaussian aspects of thermal and kinetic Sunyaev-Zel'dovich
Effects}
\author{Asantha Cooray}
\address{
Department of Astronomy and Astrophysics, University of Chicago,
Chicago, IL 60637\\
E-mail: asante@hyde.uchicago.edu}

\date{To be submitted to Phys. Rev. D. --- April 2001}

\maketitle

%------------------------------------------------------------------------------

\begin{abstract}
We discuss non-Gaussian effects associated with the local large-scale
structure contributions to the Cosmic Microwave Background  (CMB) anisotropies
through the thermal Sunyaev-Zel'dovich (SZ) effect.
The non-Gaussianities associated with the SZ effect arise from the existence
of a significant  four-point correlation function in large scale pressure fluctuations.
Using the pressure trispectrum calculated under the recently popular halo model,
we discuss the full covariance of the SZ thermal power spectrum. We
use this full covariance matrix to study the astrophysical uses of
the SZ effect and discuss the extent to which gas properties can be
derived from a measurement of the SZ power spectrum.  
With the SZ thermal effect separated in temperature fluctuations using
its frequency information, the kinetic SZ effect, also known as the
Ostriker-Vishniac effect, is expected to dominate the CMB temperature
fluctuations at small angular scales. This effect arises from the
baryon modulation of the first order Doppler effect resulting from the
relative motion of scatterers. The presence of
the SZ kinetic effect 
can be determined through a cross-correlation between the SZ thermal
and a CMB map at small scales. 
Since the SZ kinetic effect is second order, however, contributions to
such a cross-correlation arise to the lower order in the form of a
three-point correlation function, or a 
bispectrum in Fourier space. We suggest an additional statistic
that can be used to study the correlation between pressure traced by
the SZ thermal effect and the baryons traced by the SZ kinetic effect 
involving  the cross-power spectrum constructed through
squared temperatures  instead of the usual
temperature itself. Through a signal-to-noise
calculation, we show that future small angular scale multi-frequency
CMB experiments, sensitive to multipoles of a few thousand,
will be able to measure the cross-correlation of
SZ thermal and SZ kinetic effect through a temperature squared 
power spectrum.
\end{abstract}
\vskip 0.5truecm

]

%------------------------------------------------------------------------------
% User-supplied List of keywords.

%\pacs{PACS numbers: 98.80.Es,95.85.Nv,98.35.Ce,98.70.Vc
%\hfill}
%]

%------------------------------------------------------------------------------

%\tableofcontents

\vspace{1in}
Presented as part of a dissertation to the Department
of Astronomy and Astrophysics, The University of Chicago, in
partial fulfillment of the requirements for the Ph.D.

\section{Introduction}

In recent years, motivated by the advances in the experimental front,
increasing attention has been given to theoretical details related to secondary
anisotropies in the cosmic microwave background (CMB) resulting from
the low redshift large scale structure.  In addition to the direct detection,
some of these effects have also been studied in detail due to
their non-linear behavior leading to higher order correlations 
in the CMB temperature fluctuations \cite{SpeGol99,CooHu00}.
In general, the secondary anisotropies resulting from low redshifts ($z
\lesssim$ few tens) can be divided into those that
result from gravitational effects and those resulting from
scattering via free electrons in the reionized epoch. 

Through the imparted differential gravitational redshift of CMB
photons, when traversing time-evolving potential fluctuations,
the integrated Sachs-Wolfe (ISW; \cite{SacWol67}) 
contributes at large angular scales. This effect is an important
contributor at $z \lesssim 1$ in a $\Lambda$CDM universe. 
Through angular deflections
across the sky via foreground gravitational 
lensing potentials, weak gravitational lensing modifies the
CMB power spectrum at large angular scales while creating  power 
primarily at small angular scales (e.g., \cite{Sel96b,Hu00b,Zal00}
and references therein).
This lensing effect also correlates with potentials traced by other
secondary effects and produces a bispectrum in CMB temperature
data (e.g., \cite{SpeGol99,CooHu00}).

In terms of the scattering effects, the inverse-Compton scattering of CMB
photons via hot electrons, the so-called Sunyaev-Zel'dovich (SZ; \cite{SunZel80}) effect,
dominates the small angular scale signal. 
The SZ effect has now been directly imaged
towards massive galaxy clusters (e.g., \cite{Caretal96,Jonetal93}),
where the temperature of the scattering medium  can reach as high as
10 keV, producing temperature changes in the CMB of order 1 mK at
Rayleigh-Jeans wavelengths.  Given that the SZ effect also bears a
spectral  signature that differs from true temperature fluctuations,
SZ contribution can be separated in
multifrequency data, thereby suggesting a means of studying its
properties. As discussed in detail in \cite{Cooetal00a},
a multi-frequency approach can be applied to current Boomerang 
\cite{deBetal00}, and upcoming
MAP\footnote{http://map.nasa.gsfc.gov} and Planck
surveyor\footnote{http://astro.estec.esa.nl/Planck/; also, ESA
D/SCI(6)3.} missions.

The improving capabilities of hydrodynamical simulations and
analytical models involving the large scale gas distribution
have now provided predictions for the
SZ effect, mainly the SZ power spectrum and the  expected number
counts of SZ halos (see, \cite{Cooetal00a,daS99,Seletal00,Spretal00,RefTey01,Coo00}
among various other studies).
A wide-field SZ image, allowing detailed statistical studies such as
the angular power spectrum, is yet to be produced, though several
experimental attempts are currently in progress for this purpose.
These experiments include the
interferometric survey at the combined
BIMA/OVRO array (CARMA; John Carlstrom, private communication), 
the MINT interferometer (Lyman Page, private communication),
and the BOLOCAM array on the Caltech Submillimeter Observatory (Andrew
Lange, private communication).

In the present paper, we further discuss the SZ effect and address
what astrophysical properties can be deduced with a measurement of the
SZ power spectrum. For this, we require detailed knowledge of the
covariance of the SZ power spectrum beyond the simple Gaussian
sample variance. Given that the SZ effect probes the
projected pressure distribution in the local universe, its statistical
properties reflect those of pressure. As discussed in detail in Cooray
\cite{Coo00}, the statistics of large scale structure pressure
is highly non-Gaussian due to the associated non-linearities. 
The same non-Gaussianities lead to a four-point
correlation function in pressure, which in turn, 
contributes non-negligibly to the power spectrum covariance of the SZ effect.

In order to calculate the covariance associated with SZ power spectrum
measurements, we extend the semi-analytical model presented in Cooray
\cite{Coo00}  and calculate the pressure trispectrum. The full
covariance  matrix now allows us
to quantify the astrophysical abilities of SZ measurements as a probe
of gas and its temperature properties. Previous to this study, we were
unable to perform a similar calculation on how well the SZ effect 
probes gas and temperature properties due to the 
unknown covariance associated with the effect.  Using a Fisher matrix
approach, we quantify to what extent the SZ power spectrum measurements
allow determination of various properties associated with
gas and temperature evolution.

Extending our calculation on the contribution of large scale
structure gas distribution to CMB anisotropies through SZ effect, 
we also study an associated effect involving baryons 
associated with halos in the large scale structure. 
It is well known that the peculiar velocity of galaxy clusters, along
the line of sight, also lead to a contribution to temperature
anisotropies. This effect is commonly known as the kinetic, or kinematic,
Sunyaev-Zel'dovich effect and arises from the baryon density modulation of
the Doppler effect associated with the velocity field \cite{SunZel80}.
Given that both density and velocity fields are involved,
the kinetic SZ effect is essentially second order in density
fluctuations;
the thermal SZ effect is also second order beacuse of density and
temperature dependence.
Though the kinetic SZ effect was first described in \cite{SunZel80}
using massive galaxy clusters,  the same effect has been introduced under a
different context by Ostriker and Vishniac (OV; \cite{OstVis86}).
The kinetic SZ effect can be considered as the OV effect extended to
the non-linear regime of baryon fluctuations, however,
it should be understood that the basic physical
mechanism responsible for the two effects is the same.
For the purpose of this presentation, we will treat both OV effect and
the SZ kinetic effect as one contribution, though it may be easier to
think of OV as the linear contribution while kinetic SZ,
extending to non-linear regime will contain the total contribution.
Such a description has been provided by Hu in \cite{Hu00a}.
 
We calculate the kinetic SZ/OV effect, hereafter simply referred to as
the kinetic SZ effect, using the model we developed to
study the thermal SZ effect. We further extend this calculation to
consider the correlation between SZ thermal and SZ kinetic effects.
Since there is no first order cross-correlation, the lowest order
contribution to the correlation comes from a three-point function, or
a bispectrum. Cooray \& Hu discusses this bispectrum in \cite{CooHu00}.
Here, we consider an additional possibility to measure the SZ
thermal-kinetic cross-correlation via a two-point correlation function
which involves squares of the temperature instead of the usual
temperature itself. The power spectrum of squared temperatures probes
one aspect of the trispectrum resulting from the pressure-baryon
cross-correlation. Here, we show that 
there is adequate signal-to-noise for a reliable measurement
of the SZ thermal-SZ kinetic squared power spectrum measurement in
upcoming small angular scale experiments.

The layout of the paper is as follows.
In \S~\ref{sec:general}, we review the background material relevant
for current calculations on SZ thermal and SZ kinetic effects
in the context of the adiabatic cold dark matter (CDM) models and
using clustered dark matter halos as a way to calculate non-linear
clustering properties. In \S~3, we discuss the pressure trispectrum
and the covariance of pressure fluctuations in the local universe.
In \S~4, we detail the calculation of SZ effect due to large scale
structure gas distribution and calculate its trispectrum. In the same
section, using a Fisher matrix formulation, we use the
full covariance, including non-Gaussianities,
 of binned SZ power spectrum measurements to establish the
astrophysical uses of the SZ effect.
The kinetic SZ effect is discussed in \S~5, and we study the
cross-correlation between SZ thermal and SZ kinetic effects in the
form of a bispectrum and the power spectrum of squared temperatures in \S~6.
We conclude in \S~7 with a summary.

\section{General Derivation}
\label{sec:general}

We first review the properties of adiabatic CDM models relevant to
the present calculations. We then discuss the general 
properties of the halo model as applied to the calculation of the
non-linear dark matter, baryon and pressure density field power
spectra of the local large scale structure. 

\subsection{Adiabatic CDM Model}
The expansion rate for adiabatic CDM cosmological models with a
cosmological constant is
\begin{equation}
H^2 = H_0^2 \left[ \Omega_m(1+z)^3 + \Omega_K (1+z)^2
              +\Omega_\Lambda \right]\,,
\end{equation}
where $H_0$ can be written as the inverse
Hubble distance today $H_0^{-1} = 2997.9h^{-1} $Mpc.
We follow the conventions that 
in units of the critical density $3H_0^2/8\pi G$,
the contribution of each component is denoted $\Omega_i$,
$i=c$ for the CDM, $g$ for the baryons, $\Lambda$ for the cosmological
constant. We also define the 
auxiliary quantities $\Omega_m=\Omega_c+\Omega_g$ and
$\Omega_K=1-\sum_i \Omega_i$, which represent the matter density and
the contribution of spatial curvature to the expansion rate
respectively.

Convenient measures of distance and time include the conformal
distance (or lookback time) from the observer
at redshift $z=0$
\begin{equation}
\rad(z) = \int_0^z {dz' \over H(z')} \,,
\end{equation}
and the analogous angular diameter distance
\begin{equation}
\da = H_0^{-1} \Omega_K^{-1/2} \sinh (H_0 \Omega_K^{1/2} \rad)\,.
\end{equation}
Note that as $\Omega_K \rightarrow 0$, $\da \rightarrow \rad$
and we define $\rad(z=\infty)=\rad_0$.

The adiabatic CDM model possesses a two, three and four-point
correlations of the dark matter density field as defined in the usual way
\begin{eqnarray}
\left< \delta(\veck_1) \delta(\veck_2) \right>& = &(2\pi)^3
\delta_\dirac (\veck_{12}) P(k_1) \, , \\
\left< \delta(\veck_1) \delta(\veck_2)\delta(\veck_3)\right> &=&
(2\pi)^3 \delta_\dirac (\veck_{123})
B(\veck_1,\veck_2,\veck_3) \, , \\
\left< \delta(\veck_1) \ldots \delta(\veck_4)\right>_c &=&
(2\pi)^3 \delta_\dirac (\veck_{1234})
T(\veck_1,\veck_2,\veck_3,\veck_4) \, ,
\end{eqnarray}
where $\veck_{i\ldots j} = \veck_i + \ldots + \veck_j$ and
$\delta_\dirac$ is
the delta function not to be confused with the density perturbation.
Note that
the subscript $c$ denotes the connected piece, i.e. the
trispectrum is defined to be identically zero for a Gaussian field.
Here and throughout, we occasionally suppress the redshift dependence
where no confusion will arise.

In linear perturbation theory\footnote{It should be understood that
``$\lin$'' denotes here the
lowest non-vanishing order of perturbation theory for the object in
question.
For the power spectrum, this is linear perturbation theory; for the
bispectrum,\
 this is second order perturbation theory, etc.},
\begin{equation}
\frac{k^3P^\lin(k)}{2\pi^2}|_{z=0} = \delta_H^2 \left({k \over
H_0} \right)^{n+3}T^2(k) \, .
\end{equation}
We use the fitting formulae of Eisenstein \& Hu in \cite{EisHu99} to evaluate the
transfer function $T(k)$ for CDM models.
Here, $\delta_H$ is the amplitude of present-day density fluctuations
at the Hubble scale.

The bispectrum in perturbation theory is given by
\footnote{The
kernels $F_n^{\rm s}$ are derived in \cite{Goretal86} (see,
equations A2 and A3 of \cite{Goretal86}; note that their
$P_n\equiv F_n$), and we have written such that the symmetric form of
$F_n$'s are used. The use of the symmetric form accounts for the
factor of 2 in Eqs.~(\ref{eqn:bpt}) and factors of 4 and 6 in
(\ref{eqn:tript}).}
\begin{eqnarray}
B^\lin(\veck_p,\veck_q,\veck_r) &=& 2 F_2^{\rm
s}(\veck_p,\veck_q)P(k_p)P(k_q)
+ 2\; {\rm Perm.} \, , \nonumber \\
\label{eqn:bpt}
\end{eqnarray}
with $F_2^{\rm s}$ term given by second order gravitational
perturbation calculations.

Similarly, the perturbation theory trispectrum follows from Fry \cite{Fry84}:
\begin{eqnarray}
&& T^\lin =6 \left[F_3^{\rm s}(\veck_1,\veck_2,\veck_3)P(k_1)P(k_2)P(k_3) + {\rm Perm.}\right]\nonumber\\
&+& 4\left[F_2^{\rm s}(\veck_{12},-\veck_1) F_2^{\rm s}
	(\veck_{12},\veck_3)P(k_1)P(k_{12})P(k_3)+{\rm Perm.}\right] \, . 
\label{eqn:tript}
\end{eqnarray}
The permutations involve a total
of 4 terms in the first set and 12 terms in the second set.

In linear theory, the density field may be scaled backwards to higher redshift
by the use of the growth function $G(z)$, where
$\delta(k,r)=G(r)\delta(k,0)$ \cite{Pee80}
\begin{equation}
G(r) \propto {H(r) \over H_0} \int_{z(r)}^\infty dz' (1+z') \left(
{H_0
\over H(z')} \right)^3\,.
\end{equation}
Note that in the matter dominated epoch $G \propto a=(1+z)^{-1}$.

The continuity equation relates the density and velocity
fields, in the linear regime, via,
\begin{eqnarray}
{\bf v} =  -i \dot G \delta(k,0){ {\bf k} \over k^2 }\,,
\label{eqn:continuity}
\end{eqnarray}
where overdots represent derivatives with respect to radial distance
$\rad$. In this paper, we ignore the contribution to the velocity
field within individual non-linear structures, including a curl
component, and only consider this potential bulk flow. This large
scale velocity field is independent of the non-linear density
fluctuations, and thus, there is no correlation between this bulk flow
and the density field within halos.

For fluctuation spectra and growth rates of interest here,
reionization of the universe is expected to occur rather late $z_\ri
\lesssim 50$  such that the reionized media is optically thin to Thomson scattering of CMB photons
$\tau \lesssim 1$.
The probability of last scattering within $d \rad$ of $\rad$ (the visibility function) is
\begin{equation}
g =  \dot \tau e^{-\tau} = X H_0 \tau_H (1+z)^2 e^{-\tau}\,.
\end{equation}
Here 
$\tau(r) = \int_0^{\rad} d\rad \dot\tau$ is the optical depth out to
$r$,
$X$ is the ionization fraction,
\begin{equation}
       \tau_H = 0.0691 (1-Y_p)\Omega_g h\,,
\end{equation}
is the optical depth to Thomson
scattering to the Hubble distance today, assuming full
hydrogen ionization with
primordial helium fraction of $Y_p$. 
Note that the ionization
fraction can exceed unity:
$X=(1-3Y_p/4)/(1-Y_p)$  for singly ionized helium,
$X=(1-Y_p/2)/(1-Y_p)$ for fully ionized helium.  

Although we maintain generality in all derivations, we
illustrate our results with the currently favored $\Lambda$CDM
cosmological model. The parameters for this model
are $\Omega_c=0.30$, $\Omega_g=0.05$, $\Omega_\Lambda=0.65$, $h=0.65$,
$Y_p = 0.24$, $n=1$, $X=1$, with a normalization such that
mass fluctuations on the $8 h$ Mpc$^{-1}$
scale is  $\sigma_8=0.9$, consistent with observations on the
abundance of galaxy clusters \cite{ViaLid99} 
and COBE normalization \cite{BunWhi97}.  A reasonable value here
is important since  higher order correlations is nonlinearly dependent on the
amplitude of the density field.

\begin{figure}[t]
\centerline{\psfig{file=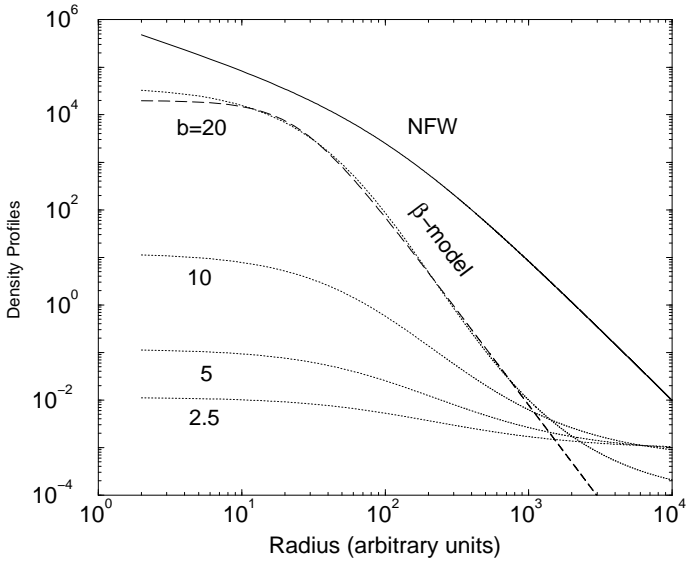,width=3.6in,angle=-90}}
\caption{The dark matter (NFW) profile and the ones predicted 
by the hydrostatic equilibrium for gas, as a function of the $b$
parameter (see, Eq.~\ref{eqn:b}) with $r_s=100$. For comparison, we also show a
typical example of the so-called $\beta$ model $(1+r^2/r_c^2)^{-3\beta/2}$
which is generally used as a fitting function for X-ray and SZ
observations of clusters. }
\label{fig:profiles}
\end{figure}

For the Sunyaev-Zel'dovich effects discussed here, we are more
interested in the clustering properties of pressure and baryons, 
rather than the
dark matter density field. We do not have a reliable way to calculate
the pressure, baryon power spectra  and their higher order correlations
and cross-correlations analytically. 
We will introduce the semi-analytic halo model for this
purpose following \cite{Coo00}. 

\subsection{Halo Approach}
\label{sec:halomodel}

In order to calculate the non-linear clustering of the dark matter,
baryon and pressure density fields, we use the halo model
described in \cite{Sel00,CooHu01a,Cooetal00b}.

Underlying the halo approach is the assertion that dark matter halos
of virial mass $M$ are
locally biased tracers of density perturbations in the linear regime.
In this case, functional relationship between the over-density of halos and
mass can be expanded in a Taylor series
\begin{eqnarray}
&&\delta_\hal({\bf x},M;z) = \nonumber \\ 
&&b_0 +
b_1(M;z)\delta_\lin({\bf x};z)   + {1 \over 2} b_2(M;z)\delta^2_\lin({\bf x};z) + \ldots
\label{eqn:bias}
\end{eqnarray}

The over-density of halos can be related to more familiar mass function
and the halo density profile by noting that we can model the fully
non-linear density field as a set of correlated discrete objects or
halos with profiles
\begin{equation}
\rho(\bfx) = \sum_{i} \rho_\halo (\bfx -  \bfx_i;M_i) \, ,
\end{equation}
where the sum is over all halo positions. 
The density fluctuation in Fourier space is
\begin{eqnarray}
\delta(\veck)& = & \sum_i e^{i \veck \cdot \bfx_i}
\delta_{\halo}(\veck,	M_i) \,.
\end{eqnarray}

Following Peebles in \cite{Pee80}, we divide space into
sufficiently
small volumes $\delta V$ that they contain only one or zero halos of a
given mass and convert the sum over halos to a sum over the
volume elements and masses 
\begin{eqnarray}
\delta(\veck)
             & = & \sum_{V_1,M_1}
                        n_1 e^{i \veck \cdot \bfx_1}
\delta_\halo(\veck,M_1)\,.
\label{eqn:fourierdelta}
\end{eqnarray}
By virtue of the small volume element
$n_1=n_1^2=n_1^\mu = $ $1$ or $0$ following \cite{Pee80}.

Using the fact that halos are biased tracers of the density field such
that their number density fluctuates as
\begin{eqnarray}
&&{dn \over d M}(\bfx;z) = {d\bar n \over d M}(M;z) \nonumber \\
&\times&        [b_0 + b_1(M;z) \delta_\lin(\bfx;z) + {1 \over 2} b_2(M;z)
\delta_\lin^2(\bfx;z)\ldots]\, , \nonumber \\
\label{eqn:fluctuate}
\end{eqnarray}
we can write
\begin{eqnarray}
\left< n_1 \right> &=& { d\bar n \over dM_1} \delta M_1 
\, ,\\
\left< n_1 n_2 \right> &=&
        \left<n_1\right>^2 \delta_{12} +
        \left<n_1\right> \left<n_2 \right>
        [b_0^2 + b_1(M_1)b_1(M_2) \, ,
        \nonumber\\
        &&
                \times \left< \delta_\lin(\bfx_1)\delta_\lin
        (\bfx_2) \right>] \, . \nonumber\\
\left< n_1 n_2 n_3 \right> &=& \ldots \, .
\label{eqn:expectation}
\end{eqnarray}
In Eq.~\ref{eqn:fluctuate}, $b_0 \equiv 1$ and $\delta_{12}$ is the Dirac
delta function and we have included only the lowest order terms.
The halo bias parameters are given in \cite{Moetal97}:
\begin{equation}
b_1(M;z) = 1 + \frac{\nu^2(M;z) - 1}{\delta_c} \, ,
\end{equation}
and
\begin{eqnarray}
b_2(M;z) &=& \frac{8}{21}[b_1(M;z)-1] + { \nu^2(M;z) -3 \over
\sigma^2(M;z)}\,.
\end{eqnarray}
Here, $\nu(M;z) = \delta_c/\sigma(M;z)$ and
$\sigma(M;z)$ is the rms fluctuation within a top-hat filter at the
virial radius corresponding to mass $M$,
and $\delta_c$ is the threshold over-density of spherical
collapse (see \cite{Hen00} for useful fitting functions). 

The derivation of the higher point functions in Fourier space is
now a straightforward but tedious exercise in algebra (see,
\cite{Coo01}).   The Fourier
transforms inherent in eqn.~(\ref{eqn:fourierdelta}) convert the correlation
functions in eqn.~(\ref{eqn:expectation}) into the power spectrum,
bispectrum, trispectrum, etc., of perturbation theory. 

Following \cite{CooHu01a}  and \cite{Coo00}, it is now convenient to define a
general integral over the halo mass function $d\bar{n}/dM$.
Though we presented the description of halo
clustering for dark matter, we can generalize this discussion to
consider any physical property associated with halo; one simply
relates the over-density of halos through the density profile
corresponding to the property of interest in Eq.~\ref{eqn:bias}.
Since we will encounter dark matter, pressure and baryon density
fields through out this paper, we write a general integral that
applies to all these three properties 
\begin{eqnarray}
&& I_{\mu,i_1\ldots i_\mu}^{\beta,\eta}(k_1,\ldots,k_\mu;z)
\equiv
\int dM  \frac{d\bar{n}}{dM}(M,z) b_\beta(M;z)  
\nonumber\\
&& \times 
T_e(M;z)^\eta y_{i_1}(k_1,M;z)\ldots y_{i_\mu}(k_\mu,M;z)\,.
\label{eqn:I}
\end{eqnarray}
Here, in addition to the dark matter, to account for clustering
properties of
pressure associated with baryons in large scale structure, we have
introduced the electron temperature, $T_e(M;z)$.

In Eq.~\ref{eqn:I}, the three-dimensional Fourier transform of the
density
fluctuation, through the halo profile of the density distribution
of any physical property $\rho_i(r,M;z)$, is
\begin{equation}
y_i(k,M;z) = \frac{1}{\rho_{bi}} \int_0^{r_v} dr\, 4 \pi r^2
\rho_i(r,M;z)
\frac{\sin(kr)}{kr} \, ,
\end{equation}
with the background mean density of the same quantity given by
$\rho_{bi}$.
Since in this paper we discuss the dark matter, pressure and baryons,
the index $i$ will be used to represent the density, $\delta$
(with $y \equiv y_\delta$), the baryons, $g$ (with $y \equiv y_g$),
or pressure, $\Pi$ (with $\equiv y_g$). Note that for both dark matter
density and baryon
clustering, $\eta=0$, as there is no temperature contribution, but for
clustering of pressure, $\eta=\mu$ when $i_1\ldots i_\mu$ describes
pressure; therefore, we will no longer make use of the index $\eta$ in the
following discussion. One additional note here is that the profile used for
baryons will be the same as the profile that we will use for
pressure. The only difference between baryon clustering and pressure
clustering is that we weigh the latter with the electron temperature,
leading to a selective contribution from electrons with the highest
temperature,while the former includes all baryons.

\begin{figure}[!h]
\centerline{\psfig{file=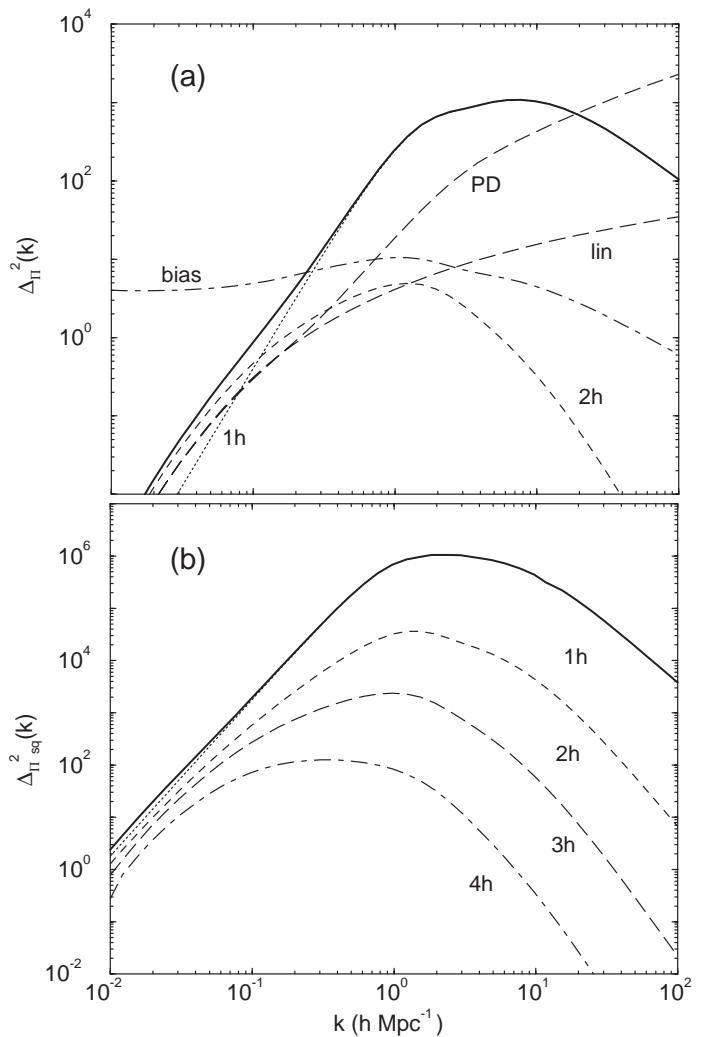,width=3.6in}}
\caption{The pressure power spectrum (a) and square-configuration
trispectrum (b) today ($z=0$)
broken into individual contributions under the halo description.
The line labeled 'bias' shows the pressure bias relative to the 
dark matter power spectrum under the halo model. In (b), we
show the square configuration trispectrum (see text).
In both (a) and (b), at all scales relevant for the Sunyaev-Zel'dovich effect,
the single halo term dominates.}
\label{fig:power}
\end{figure}

\begin{figure}[!h]
\centerline{\psfig{file=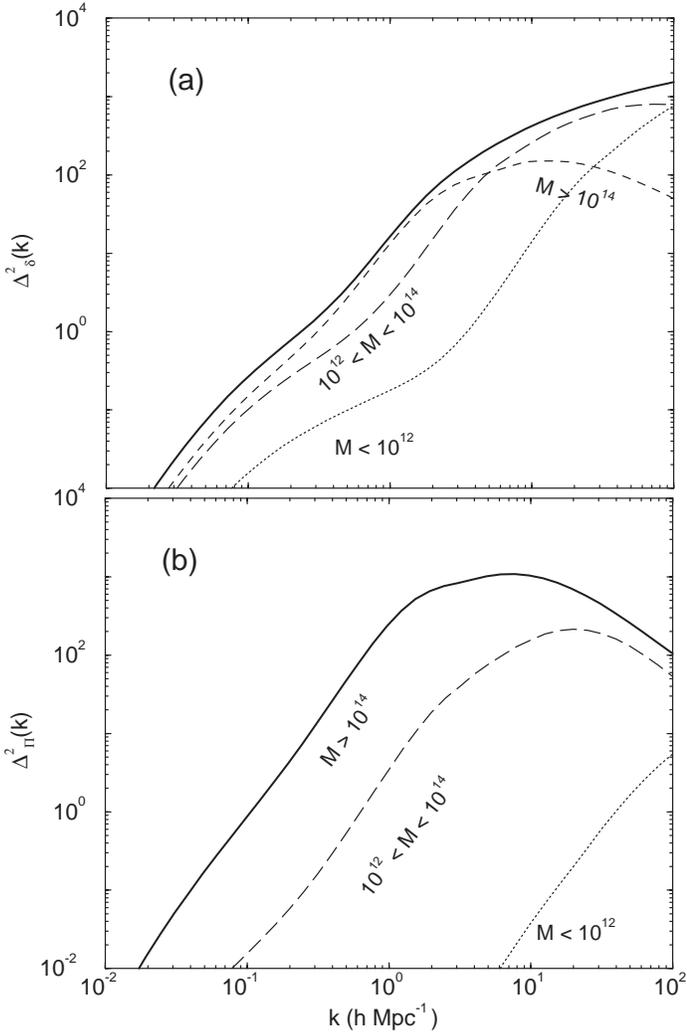,width=3.6in}}
\caption{The mass dependence on the dark matter power spectrum (a) and
pressure power spectrum (b). Here, we show the total contribution
broken in mass limits as written on the figure.
As shown in (a), the large scale contribution to the dark matter power
comes from massive halos while small mass halos contribute at small
scales. For the pressure, in (b), only massive halos above a mass of
$10^{14}$ M$_{\sun}$ contribute to the power.}
\label{fig:powermass}
\end{figure}

\subsection{Ingredients}

The dark matter profile of collapsed halos are taken to be the NFW
\cite{Navetal96} with a density distribution
\begin{equation}
\rho_\delta(r) = \frac{\rho_s}{(r/r_s)(1+r/r_s)^{2}} \, .
\end{equation}
The density profile can be integrated and related to the total dark
matter mass of the halo within $r_v$
\begin{equation}
M_\delta =  4 \pi \rho_s r_s^3 \left[ \log(1+c) - \frac{c}{1+c}\right]
\label{eqn:deltamass}
\end{equation}
where the concentration, $c$, is $r_v/r_s$.
Choosing $r_v$ as the virial radius of the halo, spherical
collapse tells us that
$M = 4 \pi r_v^3 \Delta(z) \rho_b/3$, where $\Delta(z)$ is
the over-density of collapse and $\rho_b$ is the background matter
density today. We use comoving coordinates throughout.
By equating these two expressions, one can
eliminate $\rho_s$ and describe the halo by its mass $M$ and
concentration $c$. We use the
Press-Schechter \cite{PreSch74} 
mass function to describe 
the halo mass function written as $d\bar{n}/dM(M;z)$. Following the results from $\Lambda$CDM simulations by
\cite{Buletal00}, we take a concentration-mass relationship
which is consistent with the mean: 
\begin{equation}
c(M;z)= 9(1+z)^{-1}\left[\frac{M}{M_\star(z)}\right]^{-0.13} \, ,
\label{eqn:conc}
\end{equation}
where $M_\star(z)$ is the non-linear mass scale at which the
peak-height threshold $\nu(M;z)$ is equal to one. Due to computational
limitations, we ignore the 
distribution of concentrations, for a given mass, observed in simulations by
\cite{Buletal00} and only use 
the mean value. As discussed in \cite{CooHu01b}, the
concentration distribution is important for higher moments and
ignoring the distribution in present work, it is likely that we
underestimate the non-Gaussian contribution from the single halo term
at a level of 20\%. We also expect uncertainties at the same level
due to our modeling of halos as smooth spherical distributions and by
ignoring the substructure and asphericity of halos.

The gas density profile, $\rho_g(r)$, is calculated assuming
hydrostatic equilibrium between the gas distribution and the dark
matter density field within a halo. This is a valid assumption given
that current observations of halos, mainly galaxy clusters, suggest
the existence of regularity relations, such as size-temperature (e.g.,
\cite{MohEvr97}), between physical properties of dark matter and
baryon distributions. 
Using the hydrostatic equilibrium allows us to write the
baryon density profile, $\rho_g(r)$, within halos as
\begin{equation}
\rho_g(r) = \rho_{g0} e^{-b} \left(1+\frac{r}{r_s}\right)^{br_s/r} \,
,
\label{eqn:gasprofile}
\end{equation}
where $b$ is a constant, for a given mass,
\begin{equation}
b = \frac{4 \pi G \mu m_p \rho_s r_s^2}{k_B T_e} \, ,
\label{eqn:b}
\end{equation}
with the Boltzmann constant, $k_B$ \cite{Maketal98} and
$\mu=0.59$, corresponding to a hydrogen mass
fraction of 76\%.  Equation~\ref{eqn:b} is derived
only under the assumption of hydrostatic equilibrium for the gas
distribution in a dark matter profile given by the NFW equation.
The hydrostatic equilibrium allows a basic physical assumption to
relate pressure and dark matter. Though effects such as mergers
between halos is expected to result in a violation of this assumption,
we do not consider variants since previous predictions made with the
above halo have been found to be consistent with numerical simulations
by Refregier \& Teyssier \cite{RefTey01}. 
In above, the normalization $\rho_{go}$ is determined under the assumption of
a constant gas mass fraction for halos comparable with the universal
baryon to dark matter ratio: $f_g \equiv M_g/M_\delta
=\Omega_g/\Omega_m$. When investigating astrophysical uses of the SZ
effect, we will vary this parameter and consider variations of gas fraction as
a function of mass and redshift.

The electron temperature can be calculated based on the virial
theorem or similar arguments as discussed in \cite{Coo00}.
Using the virial theorem, we can write
\begin{equation}
k_B T_e = \frac{\gamma G \mu m_p M_\delta}{3 r_v} \, ,
\end{equation}
with $\gamma=3/2$.
Since $r_v \propto M_\delta^{1/3}(1+z)^{-1}$ in
physical coordinates, $T_e \propto M^{2/3}(1+z)$.
The average density weighted temperature is
\begin{equation}
\left<T_e\right>_\delta  = \int dM\, \frac{M}{\rho_b} \frac{dn}{dM}(M,z) T_e(M,z) \,
.
\label{eqn:etemp}
\end{equation}

The total gas mass present in a dark matter halo within $r_v$ is
\begin{equation}
M_g(r_v) = 4 \pi \rho_{g0} e^{-b} r_s^3 \int_0^{c} dx \, x^2
(1+x)^{b/x} \, .
\label{eqn:gasmass}
\end{equation}

In Fig.~\ref{fig:profiles}, we show the NFW profile for the dark
matter and arbitrarily normalized gas profiles predicted by the
hydrostatic equilibrium and virial theorem for several values of $b$.
As $b$ is decreased, such that the temperature is increased, since $b
\propto 1/T_e$, the turn
over radius of the gas distribution shifts to higher radii. 
The relative normalization of the gas profiles are set with a gas fraction value of
0.1, while the NFW profile is arbitrarily normalized with $\rho_s=1$;
the gas profiles, therefore, simply scale with the same factor related
to true value of $\rho_s$.  As an
example, we also show the so-called $\beta$ model that is commonly
used to describe X-ray and SZ observations of galaxy clusters and for
the derivation purpose of the Hubble constant by combined SZ/X-ray
data. The $\beta$ model describes quite accurately the  underlying gas distribution
predicted by the gas profile in Eq.~\ref{eqn:gasprofile}, though, 
we find differences at the outermost
radii of halos. This difference can be used as a way to establish the
hydrostatic equilibrium for clusters with respect to a NFW like dark
matter distribution, though, any difference of gas
distribution at large radii may need to be accounted in the context of
possible substructure and mergers. 

A discussion on the comparison
between the gas profile used here and the $\beta$ model is available
in \cite{Maketal98}.
In addition, we refer the reader to \cite{Coo00} 
for a full detailed discussion on issues
related to modeling of pressure power spectrum using halos and
associated systematic errors. Comparisons of the halo model
predictions with numerical simulations are available in
\cite{Seletal00} and \cite{RefTey01}.
Similarly, issues related to modeling of
the dark matter clustering using halos is discussed in \cite{CooHu01a}
for the bispectrum and \cite{CooHu01b} for the trispectrum.

\subsection{Power Spectra and Trispectra}

For the calculations presented in this paper, we will
encounter the power spectrum and trispectrum involving
dark matter, pressure and baryons and their cross-correlations. 
We now write down the relevant Fourier space correlations
under the halo approach. To generalize the discussion, we will use the
index $i$ to represent the property of interest.

In general, the power spectrum of these three quantities 
under the halo model now becomes \cite{Sel00}: 
\begin{eqnarray}
P_i(k) &=& P^{1h}_i(k) +  P^{2h}_i(k) \,, \\
P^{1h}_i(k) & = & I_{2,ii}^0(k,k) \,, \\
P^{2h}_i(k) & = &\left[  I_{1,i}^1(k) \right]^2 P^\lin(k)\,,
\end{eqnarray}
where the two terms represent contributions from two points in
a single halo (1h) and points in two different halos (2h)
respectively. 

Similar to above, we can also define the cross power spectra between
two fields $i$ and $j$ as
\begin{eqnarray}
P_{ij}(k) &=& P^{1h}_{ij}(k) +  P^{2h}_{ij}(k) \,, \\
P^{1h}_{ij}(k) & = & I_{2,ij}^0(k,k) \,, \\
P^{2h}_{ij}(k) & = &I_{1,i}^1(k)I_{1,j}^1(k) P^\lin(k)\,.
\end{eqnarray}
It is also useful to define the bias of one field relative to the dark
matter density field
as \begin{equation}
{\rm bias}_{i}(k) = \sqrt{\frac{P_i(k)}{P_\delta(k)}} \, .
\end{equation}
We can also define a dimensionless
correlation coefficient between the two fields as
\begin{equation}
r_{ij}(k) = \frac{P_{ij}(k)}{\sqrt{P_i(k)P_j(k)}} \, .
\end{equation}
During the course of this paper, we will encounter, and use,
cross-power spectra such as the one involving baryon and pressure,
$P_{g\Pi}$, and dark matter and pressure, $P_{\delta\Pi}$.

As described in \cite{CooHu01b},
the contributions to the trispectrum
may be separated into those involving one to four halos
\begin{equation}
T_i = T^{1h}_i +  T^{2h}_i + T^{3h}_i + T^{4h}_i\,,
\end{equation}
where here and below the argument of the trispectrum is understood
to be $(\veck_1,\veck_2,\veck_3,\veck_4)$.
The term involving a single halo probes correlations of the property
of interest, e.g., dark matter, within that halo
\begin{equation}
T^{1h}_i =
I_{4,iiii}^0(k_1,k_2,k_3,k_4) \, ,
\end{equation}
and is independent of configuration due to the assumed
spherical symmetry for our halos.

The term involving two halos can be further broken up into two parts
\begin{equation}
T^{2h}_i = T^{2h}_{31,iiii} + T^{2h}_{22,iiii}\,,
\end{equation}
which represent taking three or two points in the first halo
\begin{eqnarray}
T^{2h}_{31,iiii} = P^\lin(k_1)I_{3,iii}^1(k_2,k_3,k_4)I_{1,i}^1(k_1) + 3\; {\rm
Perm.,} \\
T^{2h}_{22,iiii} = P^\lin(k_{12})I_{2,ii}^1(k_1,k_2)I_{2,ii}^1(k_3,k_4)+ 2\; {\rm
Perm.}
\end{eqnarray}
The permutations involve the 3 other choices of $k_i$ for the $I_{1,i}^1$
term in
the first equation and the two other pairings of the $k_i$'s for the
$I_{2,ii}^1$ terms in the second.
Here, we have defined $\veck_{12} =
\veck_1+\veck_2$; note that $k_{12}$ is the length of one of the
diagonals
in the configuration.
 
The term containing three halos can only arise with two points in one
halo and one in each of the others
\begin{eqnarray}
&&T^{3h}_i =
B^\lin(\veck_1,\veck_2,\veck_{34})I_{2,ii}^1(k_3,k_4)I_{1,i}^1(k_1)I_{1,i}^1(k_2)
\nonumber \\
&&+ P^\lin(k_1)P^\lin(k_2)I_{2,ii}^2(k_3,k_4)I_{1,i}^1(k_1)I_{1,i}^1(k_2) + 5\;
{\rm Perm.}\, ,
\end{eqnarray}
where the permutations represent the unique pairings of the $k_i$'s in
the $I_{2,ii}$ factors.  This term also depends on the configuration.
 
Finally for four halos, the contribution is
\begin{eqnarray}
T^{4h}_i &=&  I_{1,i}^1(k_1)I_{1,i}^1(k_2)I_{1,i}^1(k_3)I_{1,i}^1(k_4) \Big\{ T^\lin
        + \Big[ {I_{2,ii}^2(k_4) \over I_{1,i}^1(k_4) }
\nonumber\\ &&\quad \times
P^\lin(k_1)P^\lin(k_2) P^\lin(k_3)+ 3\; {\rm Perm.}\Big] \Big\},
\end{eqnarray}
where the permutations represent the choice of $k_i$ in the $I_{1,i}^1$'s in
the brackets.  Because of the closure condition expressed by the delta function,
the trispectrum may be viewed as a four-sided figure with sides
$\veck_i$.
It can alternately be described by the length of the four sides $k_i$
plus the diagonals.  We occasionally refer to elements of the
trispectrum
that differ by the length of the diagonals as different configurations
of the trispectrum. In the rest of the paper, we will encounter the
pressure trispectrum $T_\Pi \equiv T_{\Pi\Pi\Pi\Pi}$ and the
pressure-baryon cross trispectrum: $T_{g\Pi g\Pi}$.

\begin{table*}
\begin{center}
\caption{\label{tab:pressurecorr}}
{\sc Pressure Power Spectrum Correlations\\}
\begin{tabular}{lrrrrrrrrrrrr}
$k$   & 0.031 & 0.044 &  0.058 & 0.074 & 0.093 & 0.110 & 0.138 & 0.169
& 0.206 & 0.254 & 0.313 & 0.385 \\
\hline
0.031 &  1.000 & 0.243 & 0.337 & 0.386 & 0.398 & 0.396 & 0.385 & 0.371
& 0.359& 0.350 & 0.343 & 0.337 \\
0.044 & (0.019) & 1.000 & 0.442 & 0.546 & 0.576 & 0.578 & 0.566 &
0.547 & 0.533 & 0.521 & 0.513 & 0.507 \\
0.058 & (0.041) & (0.036) & 1.000 & 0.653 & 0.728 & 0.745 & 0.736 &
0.715 & 0.699 & 0.687 & 0.678 & 0.672 \\
0.074 & (0.065) & (0.075) & (0.062) & 1.000 & 0.807 & 0.865 & 0.868 &0.847 & 0.832 & 0.819 & 
0.811 & 0.805 \\
0.093 & (0.086) & (0.111) & (0.118) & (0.102) & 1.000 & 0.898 & 0.932 & 0.918 & 0.902 & 0.890 & 0.882 
& 0.876 \\
0.110 & (0.113) & (0.153) & (0.183) & (0.189) & (0.160) & 1.000 & 0.955 & 0.959 & 0.946 & 0.934 & 0.926 & 
0.921 \\
0.138 & (0.149) & (0.204) & (0.255) & (0.299) & (0.295)& (0.277) & 1.000 & 0.978 & 0.973 & 0.961 & 0.953 & 0.947 \\
0.169 & (0.172) & (0.238) & (0.302) & (0.368) & (0.368) & (0.433)& (0.434) & 1.000 & 0.982 & 0.972 & 0.962 & 0.956 \\
0.206 & (0.186)  & (0.261) & (0.334) & (0.334) & (0.412) & (0.541) &
(0.580) & (0.592) & 1.000 & 0.985 & 0.973 & 0.965 \\
0.254 & (0.186)  & (0.264) & (0.341) & (0.341) & (0.425) & (0.576) &
(0.693) & (0.737) & (0.748) & 1.000 & 0.986 & 0.973 \\
0.313 & (0.172) & (0.251) & (0.328) & (0.328) & (0.412) & (0.570) &
(0.698)& (0.778)& (0.839) & (0.858) & 1.000 & 0.986 \\
0.385 & (0.155) & (0.230) & (0.305) & (0.305) & (0.389) & (0.549) & (0.680) & (0.766) & (0.848) & (0.896) & (0.914) & 1.000 \\
\hline
$\sqrt{\frac{C_{ii}}{C_{ii}^{G}}}_{\rm \Pi}$ & 1.04 & 1.13& 1.29& 1.64& 1.95& 2.24& 3.58& 6.09& 9.27& 16.4& 21.2& 29.8 \\
\hline
$\sqrt{\frac{C_{ii}}{C_{ii}^{G}}}_{\rm \delta}$ & 1.00 & 1.01&
1.02& 1.03& 1.04& 1.07& 1.14& 1.23& 1.38& 1.61& 1.90& 2.26 \\
\end{tabular}
\end{center}
\footnotesize
NOTES.---%
Diagonal normalized covariance matrix of the binned pressure (upper
triangle) and dark matter density field (lower triangle) 
power spectrum with $k$ values in units of h Mpc$^{-1}$.
Lower triangle (parenthetical numbers) displays the covariance from
halo model in \cite{CooHu01b}. Final line shows
the fractional increase in the errors (root diagonal covariance) due to
non-Gaussianity as calculated under the halo model for pressure and
dark matter density field.
\end{table*} 

\begin{figure*}[t]
\centerline{\psfig{file=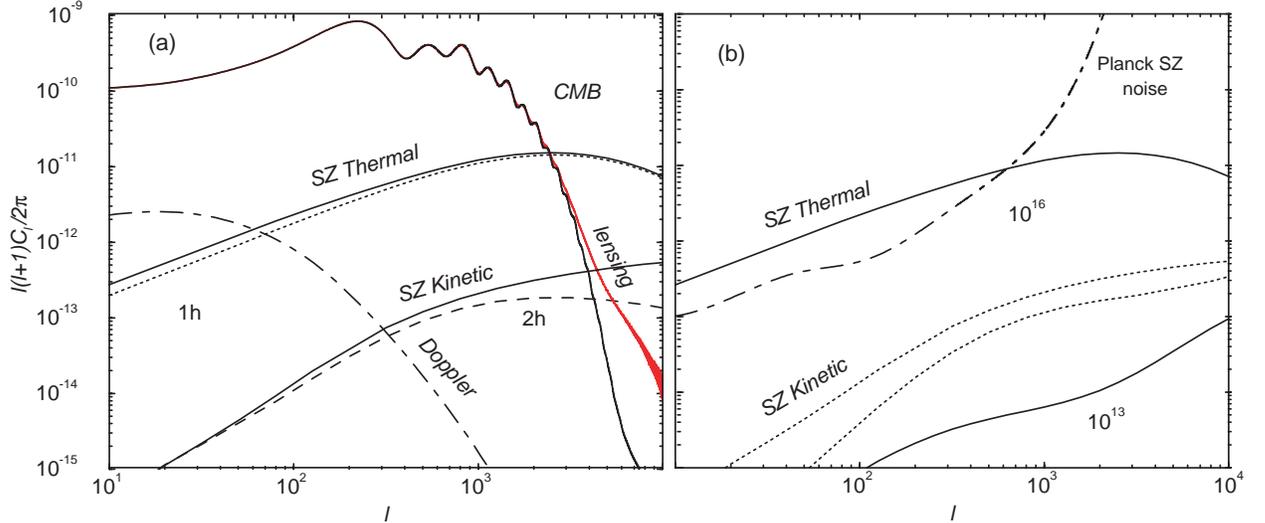,width=7in}}
\caption{The angular power spectra of SZ thermal and kinetic
effects. As shown in (a), the thermal
SZ effect is dominated by individual halos, and thus, by the single
halo term, while the kinetic effect is dominated by the large scale
structure correlations depicted by the 2-halo term. In (b), we show
the mass dependence of the SZ thermal and kinetic effects with a
maximum mass of $10^{16}$ and $10^{13}$ M$_{\sun}$. The SZ thermal
effect is strongly dependent on the maximum mass, while due to large
scale correlations, kinetic effect is not.}
\label{fig:szpower}
\end{figure*}

\section{Pressure Trispectrum and Power Spectrum Covariance}

Following \cite{MeiWhi99}, we can relate the trispectrum
to the variance of the estimator of the binned power spectrum
\begin{equation}
\hat P_i = {1 \over V } \int_{\shell i} {d^3 k \over V_{\shell i}}
\delta_\Pi^*(-\veck) \delta_\Pi(\veck)  \, ,
\end{equation}
where the integral is over a shell in $k$-space centered around $k_i$,
$V_{\shell i} \approx 4\pi k_i^2 \delta k$ is the volume of the shell
and $V$
is the volume of the survey.  Recalling that $\delta({\bf 0})
\rightarrow V/(2\pi)^3$
for a finite volume,
\begin{eqnarray}
C_{ij} &\equiv& \left< \hat P_i \hat P_j \right> -
        \left< \hat P_i \right>
        \left< \hat P_j \right>  \nonumber\\
       &=& {1 \over V} \left[ {(2\pi)^3 \over V_{\shell i} } 2 P_i^2
        \delta_{ij}+
        T_{ij}^\Pi \right]  \, ,
\end{eqnarray}
where
\begin{eqnarray}
T_{ij}^\Pi &\equiv& \int_{\shell i} {d^3 k_i \over V_{\shell i}}
        \int_{\shell j} {d^3 k_j \over V_{\shell j}}
        T_\Pi(\veck_i,-\veck_i,\veck_j,-\veck_j) \,.
\label{eqn:covarianceij}
\end{eqnarray}
Notice that though both terms
scale in the same way with the volume of the survey, only the Gaussian
piece
necessarily decreases with the volume of the shell.  For the Gaussian
piece,
the sampling error reduces to a simple root-N mode counting of
independent modes
in a shell.  The trispectrum quantifies the non-independence of the
modes both within a shell
and between shells.  Calculating the covariance matrix of the power
spectrum estimates reduces to averaging the elements of the trispectrum across
configurations
in the shell.

\subsection{Discussion}

In Fig.~\ref{fig:power}(a), we show the logarithmic
power spectrum of pressure and dark matter such that
$\Delta^2(k)=k^3 P(k)/2\pi^2$ with
contributions broken down to the $1h$ and $2h$ terms today. 
As shown, the pressure power spectrum depicts an increase in power
relative to the dark matter at scales out to few h
Mpc$^{-1}$, and a decrease thereafter. The decrease in power at small
scales can be understood through the relative contribution to pressure
as a function of the halo mass.
In Fig.~\ref{fig:powermass}, we break the total dark matter power spectrum
(a) and the total pressure power spectrum (b), to a function of mass. As
shown, contributions to dark matter come from
massive halos at large scales and by small mass halos at small scales.
The pressure power spectrum is such that through temperature weighing,
with $T_e \propto M^{2/3}$ dependence, the contribution from
low mass halos to pressure is suppressed relative to that from the high
mass end. Thus, the pressure power spectrum, at all scales of
interest, can be described with halos of mass greater than
$10^{14}$ M$_{\sun}$. A comparison of the dark matter and pressure power spectra, as a
function of mass, in Fig.~\ref{fig:powermass} reveals that the
turn-over 
in the pressure power spectrum results at an effective scale
radius for halos with mass greater than $10^{14}$ M$_{\sun}$.
As we find later, this effective turn-over produces a distinct
signature in the SZ angular power spectrum inolving a turnover of SZ
power at multipoles of few thousand; The presence of this turnover and
the exact multipole range can be used as a probe of gas and
temperature physics.

For the covariance of pressure, and also for the covariance of the SZ
power spectrum, we are mainly interested in terms of the pressure trispectrum
involving configurations that result in
$T_\Pi(\veck_1,-\veck_1,\veck_2,-\veck_2)$, i.e. parallelograms
which are defined by either the length $k_{12}$ or the angle
between $\veck_1$ and $\veck_2$.  These are the configurations that
contribute to the power spectrum covariance. For illustration purposes
we will take $k_1=k_2$ and the angle to be $90^\circ$
($\veck_2=\veck_\perp$) such that
the parallelogram is a square.
It is then convenient to define
\begin{equation}
\Delta^2_{\Pi \rm sq}(k) \equiv \frac{k^3}{2\pi^2}
T_\Pi^{1/3}(\veck,-\veck,\veck_\perp,-\veck_\perp) \, ,
\end{equation}
such that this quantity scales roughly as the logarithmic
power spectrum itself $\Delta^2(k)$.  This spectrum is
shown in Fig.~\ref{fig:power}(b) with the individual
contributions from the 1h, 2h, 3h, 4h terms shown.
As shown, almost all contributions to the pressure trispectrum come
from the single halo term.

Using the pressure trispectrum, we can now predict the pressure
covariance and, more appropriately, correlations in the binned
measurements of the pressure. The predictions made here with the halo
model to describe pressure can easily be tested in numerical simulations and the accuracy
of the halo model can be further studied. For this purpose, we
calculate the covariance matrix $C_{ij}$ from
Eqn.~(\ref{eqn:covarianceij})
with the bins centered at $k_i$ and volume $V_{\shell i} =
4\pi k_i^2 \delta {k_i}$. The binning scheme used here is the one 
we utilized in \cite{CooHu01b} to calculate the binned dark
matter power spectrum correlations. 

In Table \ref{tab:pressurecorr}, we tabulate the pressure (upper
triangle), and for comparison dark matter (lower triangle),
correlation coefficients
\begin{equation}
\hat C_{ij} = {C_{ij} \over \sqrt{C_{ii} C_{jj}}} \, .
\end{equation}
The dark matter correlations are from the halo based predictions by
 \cite{CooHu01b}. There, for the dark matter,
 we suggested that the
halo model predicted correlations agree with numerical simulations of
\cite{MeiWhi99} typically 
better than $\pm 0.1$, even in the region where
non-Gaussian effects dominate, and that the the qualitative
features such as the increase in correlations across the
non-linear scale are preserved. As we do not have measurements of the
pressure correlations from simulations, we cannot perform a detailed
comparison on the accuracy of the halo model predictions for pressure
here. 

A further test on the accuracy of the halo approach is to consider
higher order real-space moments such as the skewness and kurtosis. In
\cite{Coo00}, we discussed the SZ skewness under the
halo model. As discussed in detail by \cite{RefTey01}, halo
model predictions agree remarkably well with numerical simulations,
especially for the pressure and SZ power spectra, though, detailed
comparisons still remain to be made  with respect to bispectrum and trispectrum.

Even though the dark matter halo formalism provides a physically
motivated means of calculating the statistics of the dark matter
density field and associated properties such as pressure, 
there are several limitations of the approach that should be borne in
mind when interpreting results. The approach assumes all halos share a parameterized
spherically-symmetric profile and this assumption is
likely to affect detailed results on the configuration dependence of
the bispectrum and trispectrum.
Since we are considering a weighted average of configurations, our
predictions presented here may be insufficient to establish the
validity of the trispectrum modeling in general. Further numerical work is required to
quantify to what extent the present approach reproduces simulation
results for the full trispectrum. We do not consider such comparisons
here, other than to  suggest that the halo model has provided, at least qualitatively,
a consistent description better than any of the arguments involving a
biased description of gas tracing the dark matter etc.

\begin{figure}[t]
\centerline{\psfig{file=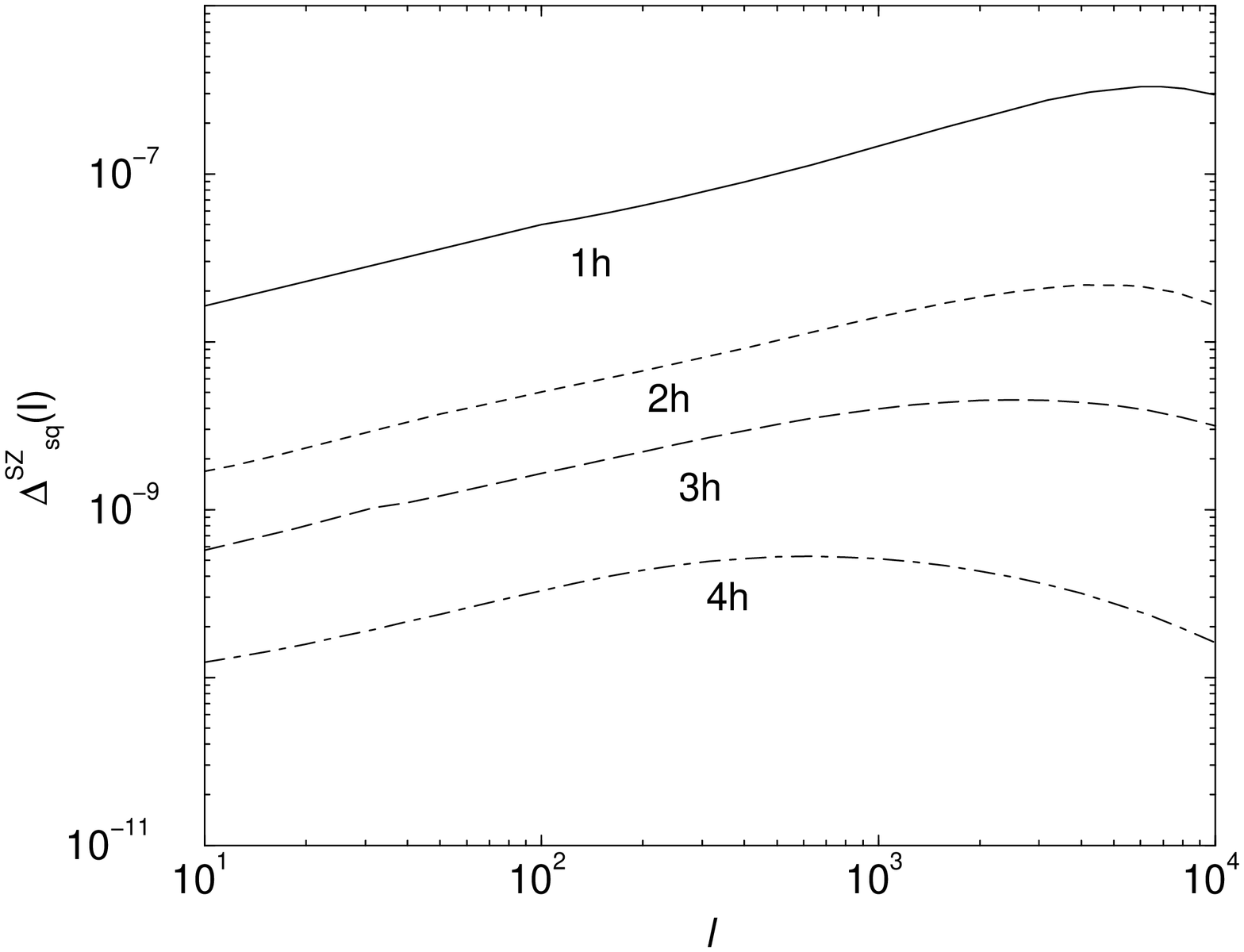,width=3.6in,angle=-90}}
\caption{SZ trispectrum for a configuration
defined by a square under
the halo model. As shown, single halo term dominates the contribution
to trispectrum at all multipoles ranging from large
angular scales to small angular scales. The dependence on the single
halo term is consistent with the general non-Gaussian behavior of the
SZ effect and it is significant non-linearity.} 
\label{fig:sztri}
\end{figure}

\section{SZ Thermal effect}

The Sunyaev-Zel'dovich (SZ; \cite{SunZel80}) effect arises 
from the  inverse-Compton scattering of CMB photons by hot electrons 
along the line of sight. 
The temperature decrement along the line of sight  due to SZ effect
can be written as the integral of pressure along the same line of
sight
\begin{equation}
y\equiv\frac{\Delta T}{T_{\rm CMB}} = g(x) \int  d\rad  a(\rad)
\frac{k_B
\sigma_T}{m_e c^2} n_e(\rad) T_e(\rad) \,
\end{equation}
where $\sigma_T$ is the Thomson cross-section, $n_e$ is the electron
number density, $\rad$ is the comoving distance, and $g(x)=x{\rm
coth}(x/2) -4$ with $x=h \nu/k_B
T_{\rm CMB}$ is the spectral shape of SZ effect. At Rayleigh-Jeans
(RJ) part of the CMB, $g(x)=-2$.
For the rest of this paper, we assume observations in the Rayleigh-Jeans
regime of the spectrum; an experiment such as Planck with sensitivity
beyond the peak of the spectrum can separate out SZ contributions
based on the spectral signature, $g(x)$ \cite{Cooetal00a}.

\subsection{Power Spectrum and Trispectrum}

The SZ power spectrum and trispectrum are
defined in the flat sky approximation
in the usual way
\begin{eqnarray}
\left< y(\bfl_1)y(\bfl_2)\right> &=&
        (2\pi)^2 \delta_\dirac(\bfl_{12}) C_l^\sz\,,\nonumber\\
\left< y(\bfl_1) \ldots
       y(\bfl_4)\right>_c &=& (2\pi)^2 \delta_\dirac(\bfl_{1234})
        T^\sz(\bfl_1,\bfl_2,\bfl_3,\bfl_4)\,.
\end{eqnarray}
These can be written as a redshift projection of the
pressure power spectrum
\begin{eqnarray}
C_l^\sz &=& \int d\rad \frac{W^\sz(\rad)^2}{d_A^2}
P_{\Pi}^\tot\left(\frac{l}{d_A},\rad\right) \, , \\
T^\sz  &=& \int d\rad \frac{W^\sz(\rad)^4}{d_A^6} T_\Pi\left(
\frac{\bfl_1}{d_A},
\frac{\bfl_2}{d_A},
\frac{\bfl_3}{d_A},
\frac{\bfl_4}{d_A},
;\rad\right) \, ,
\label{eqn:szpower}
\end{eqnarray}
where $d_A$ is the angular diameter distance. At RJ part of the
frequency spectrum,  the SZ weight function is
\begin{equation}
W^\sz(\rad) = -2 \frac{k_B \sigma_T \bar{n}_e}{a(\rad)^2 m_e c^2}
\end{equation}
where $\bar{n}_e$ is the mean electron density today. In deriving
Eq.~(\ref{eqn:szpower}),
we have used the Limber approximation \cite{Lim54} by setting
$k = l/d_A$ and flat-sky approximation. 

In Fig.~\ref{fig:szpower}(a), we show the  SZ power spectrum due to
baryons present in virialized halos.
As shown, most of the contributions to SZ power
spectrum comes from individual massive halos, while the halo-halo
correlations only contribute at a level of 10\% at large angular
scales. This is contrary to, say, the lensing convergence power
spectrum discussed in \cite{Cooetal00b}, where most of the power
at large angular scales is due to the halo-halo correlations. The
difference can be understood by noting that the SZ effect is strongly
sensitive to the most massive halos due to $T \propto M^{2/3}$
dependence in temperature and to a lesser, but somewhat related,
extent that the weight function increases towards
low redshifts. The lensing weight function selectively probes
the large scale dark matter density power spectrum at comoving
distances half to that of background sources ($z \sim 0.2$ to 0.5 when
sources are at a redshift of 1), but has no extra dependence on mass.
The fact that the SZ power spectrum results mainly from the single
halo term also results in a sharp reduction of power when the maximum
mass used in the calculation is varied. For example, as discussed in
\cite{Coo00} and illustrated in Fig~\ref{fig:szpower}(b), with
the maximum mass decreased from $10^{16}$ to $10^{13}$ M$_{\sun}$, the
SZ power spectrum is reduced by a factor nearly two orders of magnitude
in large scales and an order of magnitude at $l \sim 10^{4}$. 

\begin{figure}[t]
\centerline{\psfig{file=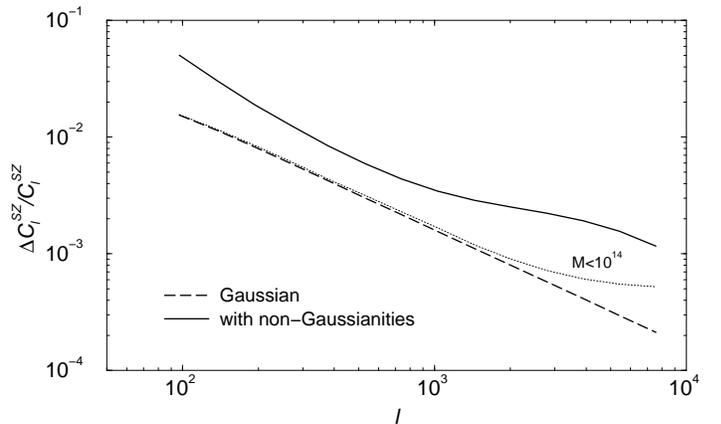,width=3.6in,angle=-90}}
\caption{The fractional errors in the measurements of the SZ
band powers. Here, we show the fractional errors under the Gaussian
approximation, and the total including non-Gaussianities. As shown, the total
contribution as a function of mass is sensitive to the presence of
most massive halos in the universe. The non-Gaussian term is
essentially dominated by the single halo term.}
\label{fig:trivariance}
\end{figure}

In Fig~\ref{fig:sztri}, we show the scaled trispectrum
\begin{equation}
\Delta^\sz_{\rm sq}(l) = \frac{l^2}{2\pi}
T^\sz(\vecl,-\vecl,\vecl_\perp,-\vecl_\perp)^{1/3} \, .
\end{equation}
where $l_\perp=l$ and $\vecl \cdot \vecl_\perp=0$.
The projected SZ trispectrum again shows the same behavior as the
pressure  trispectrum. As
shown, the contributions to the trispectrum essentially comes from the
single halo term at all multipoles. This is consistent with our
observation that SZ power spectrum is essentially dominated by the
correlations of pressure within halos. 
As discussed in \cite{Coo00}, the SZ bispectrum
is also dominated by the single halo term. Given this
dependence on the single halo term, for the
rest of the discussion involving SZ covariance, 
we will only use the single halo
term and ignore the contributions arising from large scale
correlations associated with halos.

\subsection{SZ Power Spectrum Covariance}

For the purpose of this calculation, we assume that planned small
angular scale SZ experiments will measure the thermal SZ 
power spectrum in bins of 
 logarithmic band powers at several $l_i$'s in multipole space with 
thickness $\delta l_i$. Thus, we can write the band power measurements as
\begin{equation}
\bp_i =
\int_{\shell i}
{d^2 l \over{A_{\shell i}}}
\frac{l^2}{2\pi} y(\bf l) y(-\bf l) \, ,
\end{equation}
where $A_\shell(l_i) = \int d^2 l$ is the area of 2D shell in
multipole and can be written as $A_\shell(l_i) = 2 \pi l_i \delta l_i
+ \pi (\delta l_i)^2$.

We can now write the signal covariance matrix as
\begin{eqnarray}
C_{ij} &=& {1 \over A} \left[ {(2\pi)^2 \over A_{\shell i}} 2 \bp_i^2
+ T^\sz_{ij}\right]\,, \nonumber \\
T^\sz_{ij}&=&
\int {d^2 l_i \over A_{\shell i}}
\int {d^2 l_j \over A_{\shell j}} {l_i^2 l_j^2 \over (2\pi)^2}
T^\sz(\bfl_i,-\bfl_i,\bfl_j,-\bfl_j)\,, 
\label{eqn:variance}
\end{eqnarray}
where
$A$ is the area of the survey in steradians.  Again the first
term is the Gaussian contribution to the sample variance and the
second term is the non-Gaussian contribution.
A realistic survey will also have an additional noise variance due to
the instrumental effects and a covariance resulting from the
uncertainties associated with the separation of the SZ effect from thermal
CMB and other foregrounds.

\begin{table*}
\begin{center}
\caption{\label{tab:cov}}
{\sc SZ Thermal Power Spectrum Correlations\\}
\begin{tabular}{ccccccccccccccc}
$\ell_{\rm bin}$ &
       97&   138  & 194&   271 & 378  & 529  & 739 & 1031& 1440 & 2012 & 2802 & 3905 & 5432 & 7568\\
\hline
 97 & 1.00 & 0.90 & 0.71 & 0.49& 0.30 & 0.18 & 0.09& 0.05& 0.02& 0.01 & 0.00 & 0.00 & 0.00 & 0.00 \\
138 &   (0.00) & 1.00 & 0.90 & 0.70& 0.49 & 0.30 & 0.17 & 0.09 & 0.05& 0.03 & 0.02 & 0.02 & 0.01 & 0.00 \\
194 &	(0.00) &  (0.00) & 1.00 & 0.89& 0.70 &0.48 & 0.30 & 0.17 & 0.09 & 0.05 & 0.03 & 0.03 & 0.01 & 0.00 \\
   271 &(0.00) &   (0.00) & (0.00)& 1.00& 0.89 & 0.69 & 0.47 & 0.30 & 0.17 & 0.09  &0.05 & 0.02 & 0.01 & 0.00 \\
   378 & (0.00) & (0.00) & (0.00)& (0.00)& 1.00 & 0.89 & 0.69 & 0.49 & 0.30  &0.17 & 0.09 & 0.05 & 0.03   & 0.01\\
   529 & (0.00) & (0.00) & (0.00)& (0.00)&(0.00)& 1.00 & 0.88 & 0.69 & 0.48  &0.30 & 0.17 & 0.17 & 0.11 & 0.07\\
   739 & (0.00) & (0.00) & (0.00)& (0.00)&(0.00)&(0.00)& 1.00 & 0.88 & 0.68  & 0.49 & 0.38 & 0.28 & 0.20& 0.13\\
  1031 & (0.00) & (0.00) & (0.00)& (0.00)&(0.00)&(0.00)& (0.00)& 1.00  & 0.87&0.68  & 0.56 & 0.43 & 0.31 & 0.21\\
  1440 & (0.00) & (0.00) & (0.00)& (0.00)&(0.00)&(0.00)& (0.00)& (0.00)& 1.00 &0.88 &0.69 & 0.61 & 0.45 & 0.31\\
  2012 & (0.00) & (0.00) & (0.00)& (0.00)&(0.00)&(0.00)& (0.00)& (0.01)&(0.05)&1.00& 0.88 & 0.69 & 0.60 & 0.42\\
  2802 & (0.00) & (0.00) & (0.00)& (0.00)&(0.00)&(0.00)& (0.00)&(0.02)&(0.09)&(0.39)&1.00  & 0.88  & 0.70 & 0.56\\
  3905 & (0.00) & (0.00) & (0.00)& (0.00)&(0.00)&(0.00)& (0.00)&(0.02)&(0.08)&(0.36)&(0.84)& 1.00  & 0.87 &0.70\\
  5432 & (0.00) & (0.00) & (0.00)& (0.00)&(0.00)&(0.00)& (0.00)&(0.01)&(0.06)&(0.29)&(0.65)&(0.86)& 1.00 & 0.88\\
  7568 & (0.00) & (0.00) & (0.00)& (0.00)&(0.00)&(0.00)& (0.00)&(0.01)&(0.04)&(0.20)&(0.53)    &(0.70)  & (0.88)    & 1.00  \\
\end{tabular}
\end{center}
\footnotesize
NOTES.---%
Covariance of the binned power spectrum for  the SZ effect.
Upper triangle displays the covariance found when a perfect frequency
cleaned SZ map is used to determine the SZ power spectrum.
Lower triangle (parenthetical numbers) displays the covariance found
when the variance is dominated by the primary anisotropy contribution,
as in a measurement of the SZ power spectrum in a CMB primary
fluctuations  dominated map. 
\end{table*}

\begin{figure}[t]
\centerline{\psfig{file=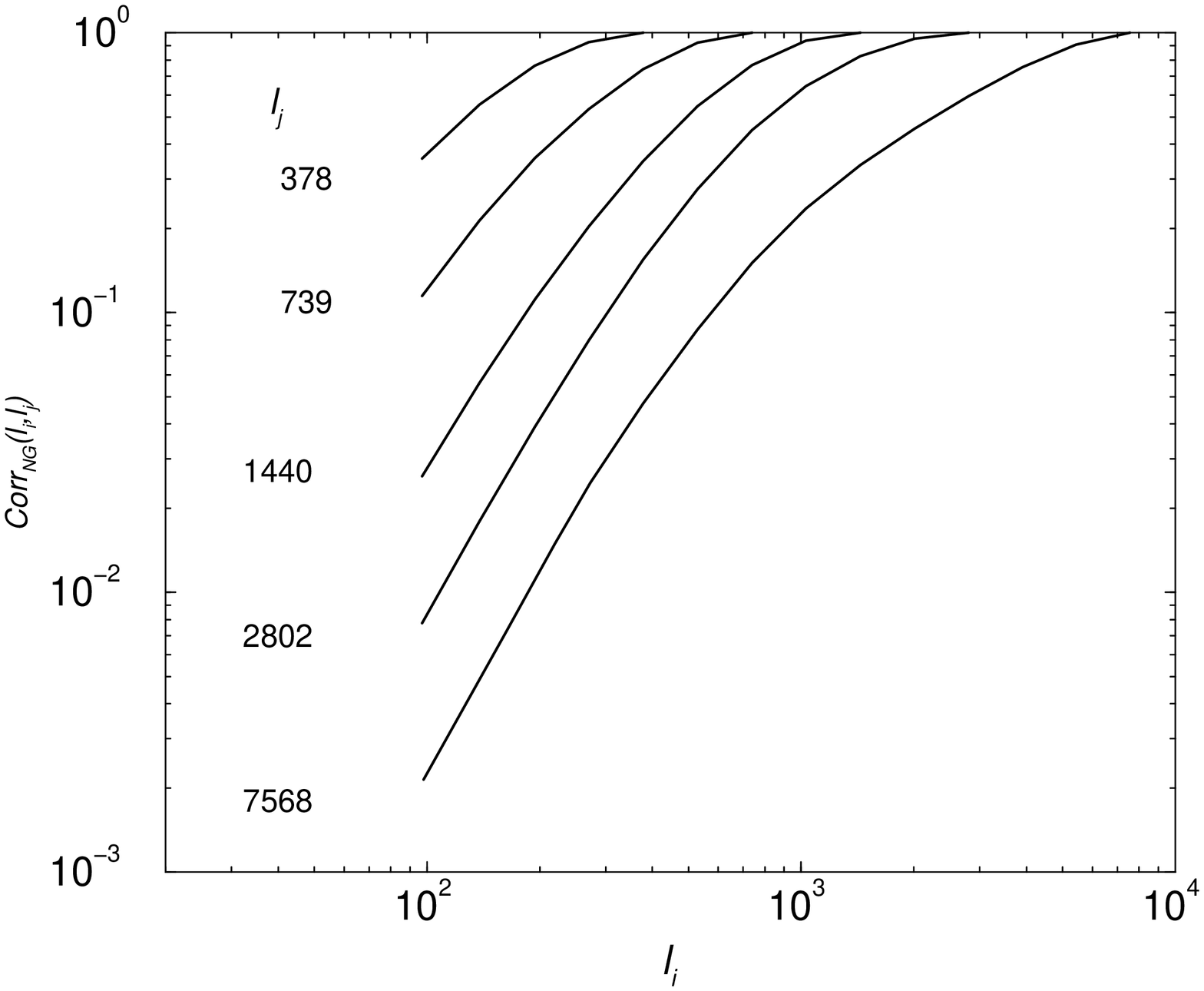,width=3.8in,angle=-90}}
\caption{The non-Gaussian correlation coefficient $\hat{C}_{ij}^\ng$,
of the SZ power spectrum, involving only the configuration of the SZ
trispectrum that contribute to the SZ power spectrum covariance (see,
Eq.~\ref{eqn:ng}). The correlations are such that they tend to 1 as
$l_i \rightarrow l_j$ and is fully described by the contribution to
the trispectrum by the single halo term.}
\label{fig:szcorr}
\end{figure}

\begin{figure}[t]
\centerline{\psfig{file=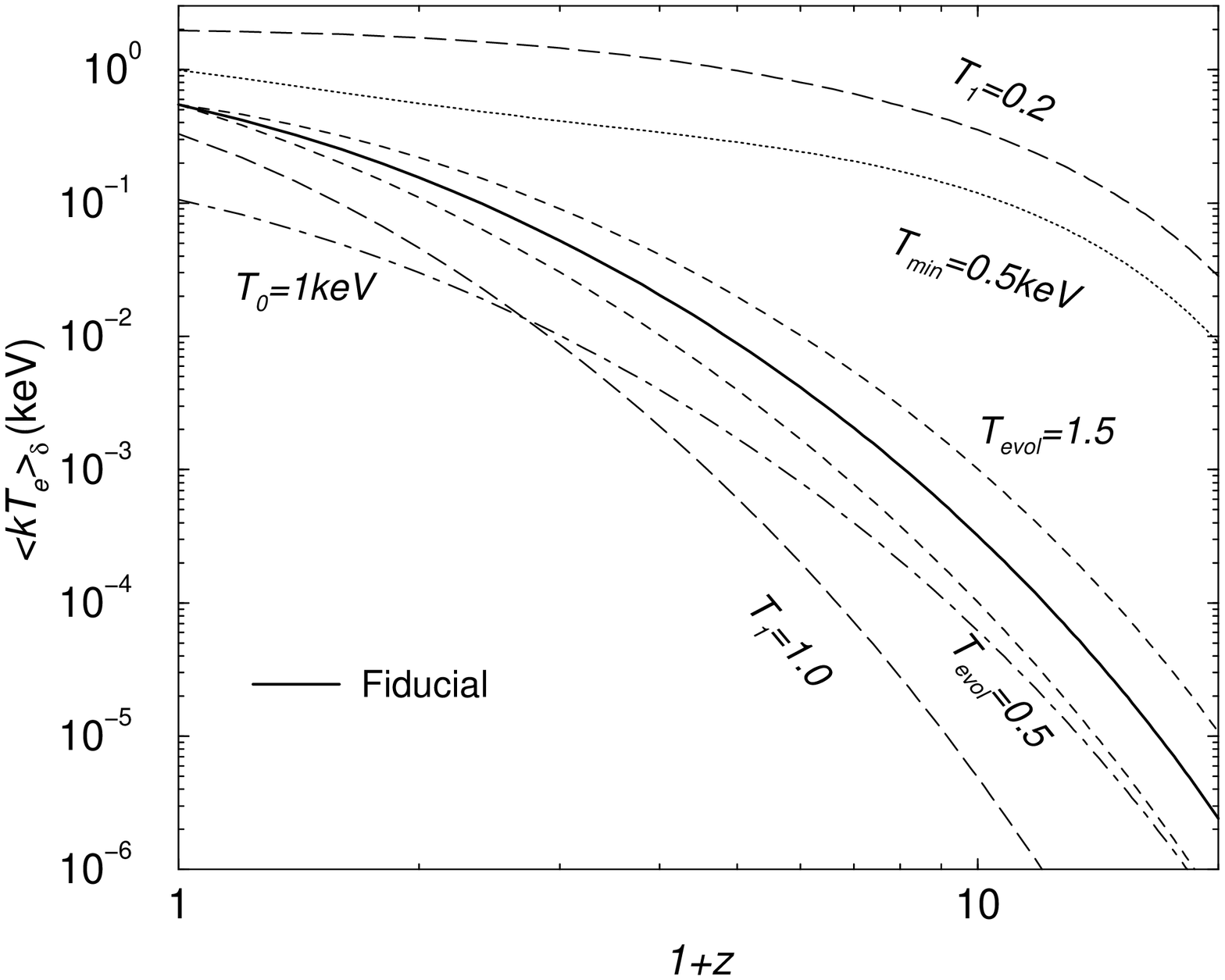,width=3.6in,angle=-90}}
\caption{The variation in the density weighted temperature  of
electrons as a function of redshift. The solid line shows the
redshift evolution of the temperature under the fiducial model while
variations about this model are shown as labeled.}
\label{fig:etemp}
\end{figure}

In Fig.~\ref{fig:trivariance}, we show the
fractional error,
\begin{equation}
{\Delta \bp_i  \over \bp_i} \equiv {\sqrt{C_{ii}}  \over \bp_i} \, ,
\end{equation}
for bands $l_i$ given in Table~\ref{tab:cov} following the
binning scheme used in \cite{CooHu01b} for the weak lensing
power spectrum.

In Fig.~\ref{fig:trivariance}, the dashed line shows the 
Gaussian error while the solid line shows the total covariance with
the addition of the SZ trispectrum (Eq.~\ref{eqn:variance}). At all
multipoles, the non-Gaussianities from the trispectrum dominates the
variance. As we discussed for the power spectrum, however, a reduction
in the maximum mass of the halos used for the SZ calculation leads to
a sharp decreases in the non-Gaussianities. With a mass cut at
10$^{14}$ $M_{\sun}$, shown by the dotted line, we see that the total
variance is consistent with the Gaussian variance out to $l \sim 1000$.

We can now write the correlation between the bands
as
\begin{equation}
\hat C_{ij} \equiv \frac{C_{ij}}{\sqrt{C_{ii} C_{jj}}} \, .
\end{equation}
In Table \ref{tab:cov} we tabulate the SZ correlations under the
assumption that the SZ power spectrum is measured independently, say
in a frequency cleaned map, (upper triangle) and is measured in the
CMB primary dominated map (lower triangle).
The correlations along individual columns increase (as one goes to
large $l$'s or small angular scales) and the maximum values are
reached at $l \sim 5000$ consistent with the general
behavior of the trispectrum.

In Fig.~\ref{fig:szcorr}, we show the non-Gaussian trispectrum
correlation coefficient given by
\begin{equation}
\hat C^{\rm NG}_{ij} =
\frac{T_{ij}}{\sqrt{T_{ii} T_{ij}}} \,.
\label{eqn:ng}
\end{equation}
As shown here,
the increase  in non-Gaussian correlation
is consistent with the fact that at all scales
it is the single halo term which dominates the non-Gaussian contribution.
Since the power spectrum is dominated by correlations in single halos,
the fixed profile of the halos correlate the power in all the
modes and the correlations between adjacent modes are significant.

The calculation, or experimental measurement,
 of the full SZ covariance is necessary for the interpretation of
observational results on the power spectrum.
The upcoming SZ surveys, where the power spectrum will be measured, is
likely to be  limited to a small area on the sky.
Thus, in the absence of many fields where the covariance can be
estimated directly from the data, the halo model based approach
suggested here provides a useful, albeit model dependent, quantification of the
covariance.  As suggested for weak lensing observations
in \cite{HuWhi00} and discussed in \cite{CooHu01b},
as a practical approach one could imagine
taking the variances estimated from the survey under
a Gaussian approximation,  after accounting for uneven
sampling and edge effects, and scaling it up by the non-Gaussian
to Gaussian variance ratio of the halo model along with
inclusion of the band power correlations. Additionally, using the
covariance as the one calculated here, one can use the approach
well known in the fields of CMB and galaxy power spectrum measurements 
to decorrelate band powers \cite{Ham97}. 

\begin{figure*}[thb]
\centerline{\psfig{file=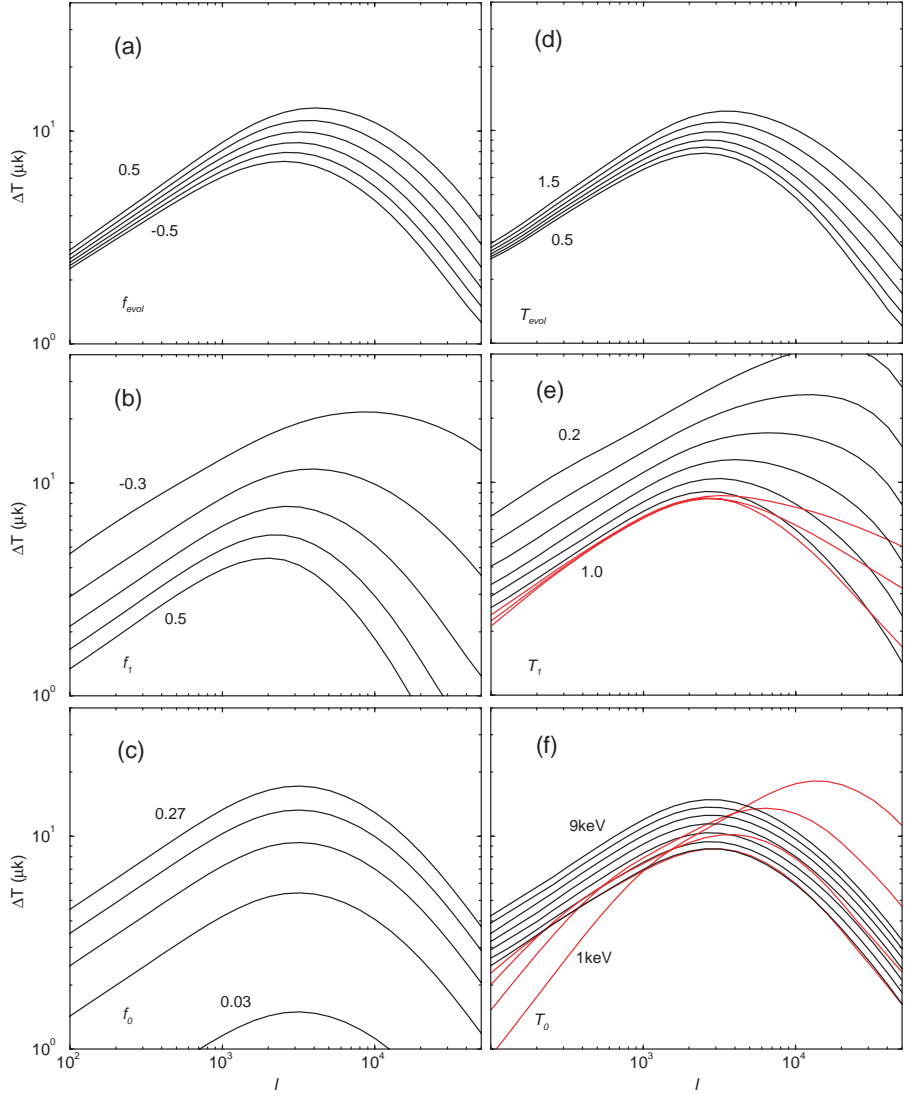,width=4.7in}}
\caption{The temperature fluctuations of the SZ effect through
variations in the astrophysical parameters under the halo model. From
(a) to (c), we show the variations associated with gas evolution while
from (d) to (f), we show variations involved with temperature. The
paraneters are described in \S~\ref{sec:parameters}.}
\label{fig:szparams}
\end{figure*}

\subsection{Astrophysical Uses of the SZ Power Spectrum}
\label{sec:parameters}

The calculation of the full covariance matrix now allows us to study
how well the SZ power spectrum measures certain astrophysical and
cosmological parameters. The upcoming CMB power spectrum
measurements, complemented by the related local universe observations
such as the galaxy power spectrum or supernovae, are expected to
constrain most of the cosmological parameters to a
reasonable accuracy \cite{Eisetal00}. Thus,
we ignore the possibility that the SZ effect can be used as
a probe of cosmology and only concentrate on the astrophysical uses of
the SZ effect. This is a reasonable approach to take since
there are many unknown astrophysics associated with the SZ effect
involving the clustering of gas density and temperature. Such an approach
allows us not to complicate the parameter measurements by adding both astrophysical
and cosmological parameters. Assuming the cosmology will be safely
known, we now ask the question what additional astrophysical 
parameters one can hope to extract from the SZ effect under the present halo model. 

There are many approaches to parameterize the unknown astrophysics of
the SZ power spectrum. Some possibilities have already been suggested
in the literature, essentially involving the gas evolution
\cite{HolCar99}.  Since the SZ effect involves both gas
and temperature as a product, ie. the pressure, one may be led to
conclude that it is not possible to separate effects associated with
temperature from those associated with gas density. Given the dependence of
temperature on the pressure profile, which we calculated from
hydrostatic equillibrium, one can expect the
 degeneracy between gas properties and 
temperature properties to be partly broken. Note that the profile
shape only depends on the temperature and not the exact density of gas;
thus, profile shape is only affected by the temperature. This
statement is exact 
under the assumption that viral equation can be applied. 
If one were to consider a varying temperature distribution, 
then the pressure profile will be sensitive
to both gas density and temperature.  We do not consider
the case with a temperature profile  as we are only interested in the most
simplest case to describe pressure here.

As discussed earlier, the clustering of pressure power spectrum has a turnover
corresponding to an equivalent scale radius of pressure.
 Through the gas pressure profile, this turnover 
can be characterized by the parameter $b$ and the dark matter scale
radius $r_s$. Note that $b \propto 1/T_e$, so a measurement of $b$
is essentially a probe of the electron temperature, though, it is
unlikely that one can obtain all information on temperature and its
evolution from one parameter measurement. Thus, instead of $b$, we
take temperature itself to be one interesting astrophysical parameter
and consider its evolution such that
\begin{equation}
T(M,z) = T_0 \left(\frac{M}{10^{15} h^{-1} M_{\sun}}\right)^{T_1}
(1+z)^{T_{\rm evol}} + T_{\rm min} \, .
\end{equation}
Here, the four parameters represent the temperature-mass
normalization, $T_0$, which in the fiducial case has the value given by the
virial equation ($K_BT_0 \sim$ 5.2 keV),
 the mass dependence slope, $T_1$, with a fiducial 
value of $2/3$, a
redshift dependent evolutionary parameter, $T_{\rm evol}$, with a 
fiducial value of 1, and a minimum temperature for gas independent of
mass and redshift $T_{\rm min}$, with a value of zero in the fiducial case. 
This latter parameter accounts for any possible preheating of gas before virializing in halos due to effects
associated with some unknown astrophysics, such as the reionization
process. A measurement of $T_{\rm min}$ would be interesting given
that observational data from massive clusters to less-massive galaxy
  groups suggest
possible preheating of gas before virialization in halos. 
As discussed
in \cite{Coo00} and \cite{Spretal00}, the SZ power
spectrum provides a strong probe of preheating of gas. 
We note here
that our description of preheating as a redshift independent temperature
is likely to be too extreme
since one expects the preheating temperature to vary with both
redshift and mass such that all three parameters, $T_0$, $T_1$ and
$T_{\rm evol}$, are affected. We do
not consider variations in $T_{\rm min}$ as a function of redshift
due to our unknown knowledge of physics associated with this process
and that addition of extra parameters can lead to further degeneracies.
In Fig.~\ref{fig:etemp}, we show the variation in the redshift evolution
of the density weighted temperature of electrons about the fiducial
model. The density weight temperature  was calculated following
Eq.~\ref{eqn:etemp}.

In addition to the temperature, the
SZ effect also depends on the number density of electrons in clusters.
So far, we have considered this number through the universal baryon
fraction in the universe such that $f_g \equiv M_g/M_\delta=\Omega_g/\Omega_m$.
This assumption ignores any possible effects associated with the
evolution of the gas fraction in halos, independent of any evolution
that may be associated with temperature. It will be interesting to study to what
extent future observations will allow the measurement of the fraction
of baryons that is responsible for the SZ effect, and any evolution
that may be associated with this fraction. Thus, a second set of
parameters one  can hope to extract from SZ observations involves 
gas mass fraction of halos and its evolution.

To study such gas properties, we parameterize the gas mass fraction such that
\begin{equation}
f_g = f_0 \left(\frac{M}{10^{15} h^{-1} M_{\sun}}\right)^{f_1}
(1+z)^{f_{\rm evol}} \, .
\end{equation}
In a recent paper, Majumdar \cite{HolCar99} has suggested the 
possible measurement of any mass and redshift dependence of gas
mass fraction in galaxy clusters,
given that the SZ power  spectrum was observed to vary 
significantly with changes in these two parameters.
Since the SZ power spectrum essentially is sensitive to 
$\sim f_g^2 T_e^2$, however,
such a suggestion for measurement of gas evolution is not
independent of any variations associated with temperature, which was
ignored in the study of \cite{HolCar99}. Our general
parameterization above involving both temperature and gas allows us to
quantify how well independent statements can be made on possible
measurement of gas density and temperature evolution, under the assumption
that cosmology is known. Note that 
gas evolution is not present in our fiducial model since we take
the gas fraction to be independent of mass and redshift 
with $f_1=0$ and $f_{\rm evol}=0$, respectively.

We now have a total of seven parameters we wish to extract from a
measurement of the SZ power spectrum. 
In order to perform this calculation  we take a  Fisher matrix based approach.
The Fisher matrix is simply a projection of the covariance
matrix onto the basis of astrophysical parameters $p_i$
\begin{equation}
{\bf F}_{\alpha\beta} = \sum_{ij}
      {\partial \bp_i \over \partial p_\alpha} (C_{\rm tot}^{-1})_{ij}
{\partial \bp_j \over \partial p_\beta} \, ,
\label{eqn:fisher}
\end{equation}
where the total covariance includes both the signal
and noise covariance.  Under the approximation of Gaussian shot
noise, this reduces to replacing $C^\sz_l \rightarrow
C^\sz_l + C^{\rm Noise}_l$ in the expressions leading up
to the covariance equation~(\ref{eqn:variance}). 
The noise power spectrum includes the noise associated with detectors,
beam size and variance resulting from the separation of the SZ effect
from other temperature fluctuations in multifrequency data.

Under the approximation that there are a sufficient number
of modes in the band powers that the distribution of power
spectrum estimates is approximately Gaussian, the Fisher matrix
quantifies the best possible errors on cosmological parameters that can
be achieved by a given survey.  In particular $F^{-1}$ is
the optimal covariance matrix of the parameters and
$(F^{-1})_{ii}^{1/2}$
is the optimal error on the $i$th parameter.
Implicit in this approximation of the Fisher matrix is the neglect of
information from the parameter dependence of the covariance matrix
of the band powers themselves. We neglect this information due to computational restrictions on the
calculation of covariance for all variations in parameters within
a reasonable amount of time. We do not expect this exclusion to change our
results significantly. Also, here, we are mostly interested  in an order
of magnitude estimate on how well the SZ power spectrum can constrain
astrophysics associated with large scale pressure.

The Fisher matrix approach allows us to
address how well degeneracies are broken in the parameter space 
under the assumption of a  fiducial model for the parameters. 
For the purpose of this calculation, we take binned measurements of
the SZ power spectrum following the binning scheme in
Table~2. We consider a perfect, no instrumental noise, 
SZ experiment with  no noise
contribution to the covariance and observations out to 
$l \sim 10^4$. To consider a real world scenario, we also study the
astrophysical uses of the SZ power spectrum that can be extracted from
the Planck mission. Here, we use the SZ noise
power spectrum calculated for Planck with detector noise and
uncertainties in the separation of SZ from CMB and other foreground
in \cite{Cooetal00a}. This SZ noise power spectrum is
shown in Fig.~\ref{fig:szpower}(b).

\begin{table}[t]
\begin{center}
\caption{\label{tab:fisher}}
{\sc Inverse Fisher Matrix ($\times$ 10$^2$)\\}
\begin{tabular}{lrrrrrrr}
$p_{i}$
        & $T_0$     & $T_1$      & $T_{\rm evol}$     & $T_{\rm min}$
& $f_0$ & $f_1$ & $f_{\rm evol}$\\
\hline
$T_0$ & 8.80 & 1.32  & 3.67 & -1.93 & -0.21 & -1.69 & -3.36\\
$T_1$&  & 0.51  & 1.08  & -0.18 & -0.04 & -0.41 & -0.07\\
$T_{\rm evol}$ & & & 2.69  & -0.62 & -0.11 & -0.94 & -2.11\\
$T_{\rm min}$ & & & & 0.48 & 0.04 & 0.29 & 0.67\\
$f_0$ & & & & & 0.006 & 0.05 & 0.09\\
$f_1$ & & & & & & 0.39 & 0.73\\
$f_{\rm evol}$ & & & & & & & 1.67\\
\end{tabular}
\end{center}
\footnotesize
NOTES.---%
Inverse Fisher matrix for the SZ effect with seven parameters and full
non-Gaussian errors. The error on an individual parameter is the
square root of the diagonal element of the inverse-Fisher matrix for the
parameter while off-diagonal entries of the inverse Fisher matrix
shows correlations, and, thus, degeneracies, between parameters. We
have assumed a perfect, no instrumental noise, experiment with a 
full sky survey ($f_\sky=1$). The seven parameters are
described in \S~\ref{sec:parameters}.
\end{table}

\subsection{Discussion}

In Fig.~\ref{fig:szparams}, we show the variation associated with SZ
temperature fluctuations $(\Delta T = \sqrt{l(l+1)/(2\pi) C_l} T_{\rm
CMB})$  for parameters involved with gas and temperature
evolutions. These plots allow us to understand some of the
degeneracies associated with the description of gas and temperature
evolution. For example, as shown,
the gas and temperature redshift dependence essentially predicts
similar behavior for the SZ temperature fluctuation, though there are
minor differences due to the temperature dependence on the pressure
profiles of halos. For the most part, variations due to temperature
evolution is due to the normalization and not due to variations in the
profile shape. In (b) and (e), we show variations due to the mass
slope of the gas evolution and temperature evolution,
respectively. Here again, we see similar behavior. When the slope of
the mass-temperature relation, as a function of mass, is greater than
0.7, we see significant differences, especially involving an increase
in temperature fluctuations at small scales. This is due to the relatively
increasing weighing of small mass halos. 

In (c) and (f), we show
variations associated with gas evolution normalization $f_0$ and
temperature-mass normalization $T_0$. The variation associated with
$f_0$ is easily understood since the effect is only a change in the
overall normalization of the power spectrum. The variation with
temperature-mass normalization shows both effects due to normalization
and the profile. When the normalization is low, gas clusters to small
radii in low mass halos leading to an increase in power at small
scales. As the temperature normalization is increased, gas profile
varies such that there is a reduction in small scale power and the
angular multipole of the turn-over scale shifts to low values. When
the temperature normalization is sufficiently high, the overall
weighing resulting from the overall temperature  multiplicative factor
becomes important. Now, the power spectrum behaves as a simple
normalization change, similar to the variation in power due to gas evolution normalization.
As shown in Fig.~\ref{fig:szparams}(a) to (f), there are significant
degeneracies involved with astrophysical parameters that lead to the
SZ effect. 

In Table~\ref{tab:errors}, we tabulate the errors on these seven
parameters using the inverse Fisher matrix for a possible SZ power
spectrum measurement. Here, we have considered the possibility that
parameter extraction will be limited to 3, 5 and 7 parameters.
The increase in number of parameters to be measured from a SZ power
spectrum increases degeneracies associated with the set of parameters
resulting in their accuracies. In the case of the 3 parameters
involving temperature-mass normalization, $T_0$, a minimum temperature for
all halos $T_{\rm min}$, and the gas mass fraction $f_0$, in a perfect
experiment, all three parameters can be extracted such that they will be provide essentially
very strong constraints. For example, the error on $f_0$  is such that one
can identify the gas fraction of clusters responsible for SZ effect 
from the cosmic mean of $\Omega_g/\Omega_m=0.05/0.3$ with an error of
$4 \times 10^{-3}$. With Planck, one can constrain the preheating temperature at the level
of $\sim$ 0.7 keV, and since current predictions for possible
preheating is also at the level of few tenths keV, Planck SZ power
spectrum can either confirm or put a useful limit on preheating
temperature at current expectations.

As tabulated, however, the accuracy to which parameters can be
determined from SZ power spectrum reduces significantly when the
number of parameters to be determined is increased. For example,
the 
Planck mission will only set a limit at $\sim$ 1.7 keV on preheating, 
if one were to study
both the  mass and redshift dependence of electron temperatures. Such an upper
limit is unlikely to be useful for current studies related to
preheating of gas. Given that we cannot obtain useful errors with
Planck for 5 parameters, we suggest that Planck may not be useful for
the purpose of studying the full parameter space suggested here. This
is understandable since Planck only allows the measurement of the SZ
power spectrum out to $l \sim$ 1500, while most of the variations due
to parameters under discussion here happens at $l \sim 5000$ or higher
where the
turnover in the SZ power spectrum is observed.  The Planck mission,
however, allows one to obtain reasonable errors on parameters which
generally define the normalization of the power spectrum, such as the
temperature-mass normalization or the normalization of gas mass fraction.
The normalization for gas mass fraction from Planck will be useful for
the purpose of understanding what fraction of cosmic baryons reside in
massive halos and contribute to the SZ effect and to look for any
discrepancy of such a value from the total baryon content
predicted by big bang nucleosynthesis arguments.
In order to obtain reliable measurements of evolution of gas and
temperature, a small scale experiment sensitive to multipoles out to
$l \sim 10^4$ will be necessary.

For  a perfect experiment, we show the errors on
seven parameters also in Table~\ref{tab:errors}. The inverse Fisher
matrix in this case is tabulated in
Table~\ref{tab:fisher}. The diagonals of the inverse Fisher matrix
show the variance of individual parameters, while, more importantly,
the off diagonals show the covariance between parameters. These
covariances allow one to understand the degeneracies between parameters.
In Table~\ref{tab:errors}, we show the full extent to which parameters degrade
the accuracies by tabulating degradation factors associated with the
seven parameters. The degradation factor list the increase in parameter error
from what can be achieved if all other parameters are known to what
can be achieved when all parameters are to be retrieved from data. 
The degradation factors are at the level of one hundred or more for some
parameters, suggesting that there are significant degeneracies
associated with the parameterization of the temperature and gas
fraction as a function of mass and redshift. Our results generally
suggest that accurate estimates 
of gas evolutionary properties, in the presence of
unknown temperature properties, are not possible. 

In addition to the parameter degeneracies, the non-Gaussianities
associated with the SZ effect also increase the errors on
parameters. For example, for the seven parameters under discussion
here and again for a perfect and full sky experiment, we list the
errors on parameters one can obtain if one were to ignore the
non-Gaussian contributions to the covariance. As tabulated,
non-Gaussianities increase the error on parameters by up to factors of
1.5, suggesting that the ignoring the non-Gaussianities will lead to a
significant underestimate of the errors in parameters. This should be
considered under the context that the SZ effect is significantly
non-Gaussian at all scales of interest and that ability to distinguish
parameters happen only at multipoles of a few thousand where the
non-Gaussianities in fact dominate.

\begin{table}[t]
\begin{center}
\caption{\label{tab:errors}}
{\sc Parameter Errors\\}
\begin{tabular}{rlrrrrrrr}
 & N     & $T_0$     & $T_1$  & $T_{\rm evol}$  & $T_{\rm min}$ & $f_0$ & $f_1$ & $f_{\rm evol}$ \\
\hline
3 & Perfect  & 0.04  &        &                 & 0.002 & 0.0004 &       &   \\
 & Planck & 0.79     &        &                 & 0.75 & 0.03  &       &   \\
\hline
5 & Perfect  & 0.13  & 0.02   & 0.06 & 0.05 & 0.002 &  & \\
 & Planck & 1.39 & 0.41 & 1.22 & 1.37 & 0.05 &  & \\
\hline
7 & Perfect & 0.30 & 0.07 & 0.17 & 0.07 & 0.008 & 0.06 & 0.13 \\
& Degradation & 47 & 184 & 133 & 34 & 82 &  240 & 130\\
& Gaussian & 0.18 & 0.04 & 0.10 & 0.04 & 0.005 & 0.04 & 0.08 \\
& Increase (\%) & 64 & 70 & 77 & 81 & 73 & 60 & 68\\
\end{tabular}
\end{center}
\footnotesize
NOTES.---%
Parameter errors, $(F^{-1})_{ii}^{1/2}$, using the halo model and the
full covariance for the SZ effect. We tabulate these errors for a
perfect experiment with no instrumental noise and full sky
observations out to $l \sim 10^4$. We also show the expected errors for
Planck mission with a useful sky fraction of 65\% ($f_\sky=0.65$),
and with the noise power spectrum shown in Fig.~\ref{fig:szpower}(b).
The parameters are described in \S~\ref{sec:parameters}.
We break the parameter estimation to consider recovery of 3, 5 and 7 parameters.
Under ``Degradation'' we tabulate the degradation factors,  $(F)_{ii}^{-1/2}/(F^{-1})_{ii}^{1/2}$, due to 
parameter degeneracies. We also list the parameter errors expected if
one were to assume Gaussian sample variance only for the SZ power
spectrum and were to ignore the non-Gaussian covariance. The increase
in error on individual parameters, with the introduction of the full
covariance matrix, ranges from 40\% to nearly 100\%.
\end{table}

\section{SZ kinetic (Ostriker-Vishniac) effect}

The bulk flow of electrons that scatter the CMB photons, in the reionized epoch, 
lead to temperature fluctuations through the well known Doppler effect:
\begin{equation}
T^\dop(\bn) = \int_0^{\rad_0} d\rad g(r) \bn \cdot {\bf v}_g(\rad,\bn
\rad)\, ,
\end{equation}
where ${\bf v}_g$ is the baryon velocity.
In Fig.~\ref{fig:szpower}(a), we show the linear Doppler effect,
including contributions resulting due to double scattering effect
described in \cite{Kai84} (see, \cite{CooHu00} for
details). The power spectrum is such that it
peaks around the horizon at the reionization projected on the sky
today. The effect cancels out significantly at scales smaller than the
horizon at scattering since photons scatter against the crests and
troughs of the perturbation. 

The Ostriker-Vishniac effect arises from the second-order
modulation of the Doppler effect by density fluctuations \cite{OstVis86},
and avoids strong
cancelation associated with the linear Doppler effect.  This nonlinear
effect is also known as the kinetic SZ effect from large-scale structure
\cite{Hu00a} and is associated with the line of sight motion of halos.
Due to the  density weighting, the kinetic SZ effect peaks at small
scales: sub arcminutes for $\Lambda$CDM.
For a fully ionized universe, contributions are broadly distributed in redshift so that the
power spectra are moderately dependent on the optical depth
$\tau$. Here, we assume an optical depth to ionization up to 0.1,
consistent with current upper limits on the reionization redshift from
CMB \cite{Grietal99} and
other observational data (see, e.g., \cite{HaiKno99} and references therein).

The kinetic SZ temperature fluctuations, denoted as $\dsz$,
 can be written as a product of the
line of sight velocity, under linear theory, and density fluctuations
\begin{eqnarray}
T^\dsz(\hat{\bf n})&=&  \int d\rad
        g(r) \hat{\bf n} \cdot {\bf v}_g(r,\bn r) \delta_g(r, \bn r)
\nonumber\\
&=&-i \int d\rad g \dot{G} G
\int \frac{d^3{\bf k}}{(2\pi)^3} \int \frac{d^3{\bf k}'}{(2\pi)^{3}}
\nonumber \\
&&\times \delta_\delta^\lin({\bf k}-{\bf k}')\delta_g({\bf k'})
e^{i{\bf k}\cdot \hat{\bf n}\rad} \left[ \hat{\bf n} \cdot 
\frac{\veck - \veck'}{|\veck - \veck'|^2}\right] \, , \nonumber \\
\end{eqnarray}
Here, we have used the linear theory to obtain the
large scale velocity field in terms of the linear dark matter density field. The
multiplication between the velocity and density fields in real space
has been converted to a convolution between the two fields in Fourier
space.

We can now expand out the temperature perturbation due to kinetic SZ
effect, $T^\dsz$, into
spherical harmonics:
\begin{eqnarray}
&&a_{lm}^\dsz = -i \int d\hat{\bf n}
\int d\rad\; (g\dot{G} G)
\int \frac{d^3{\bf k_1}}{(2\pi)^3}\int \frac{d^3{\bf
k_2}}{(2\pi)^3}\nonumber \\
&&\times \delta_\delta^\lin({\bf k_1})\delta_g({\bf k_2}) 
e^{i({\bf k_1+k_2})
\cdot \hat{\bf n}\rad} \left[ \frac{\hat{\bf n} \cdot \veck_1}{k_1^2} \right]
Y_l^{m\ast}(\hat{\bf n}) \, ,
\end{eqnarray}
where we have symmetrizised by using $\veck_1$ and $\veck_2$
to represent $\veck-\veck'$ and $\veck'$ respectively.
Using
\begin{equation}
\hat{\bf n} \cdot \veck = \sum_{m'} \frac{4\pi}{3} k
Y_1^{m'}(\hat{\bf n}) Y_1^{m'\ast}(\hat{\veck}) \, ,
\end{equation}
and the Rayleigh expansion
\begin{equation}
e^{i{\bf k}\cdot \hat{\bf n}\rad}=
4\pi\sum_{lm}i^lj_l(k\rad)Y_l^{m \ast}(\bk)
\Ylmn(\bn)\,,
\label{eqn:Rayleigh}
\end{equation}
 we can further simplify and rewrite the multipole moments as
\begin{eqnarray}
&&a_{lm}^\dsz = -i \frac{(4 \pi)^3}{3}
\int d\rad
\int \frac{d^3{\bf k}_1}{(2\pi)^3} \int \frac{d^3{\bf
k}_2}{(2\pi)^3}
\sum_{l_1 m_1}\sum_{l_2 m_2}\sum_{m'} \nonumber\\
&& \times
(i)^{l_1+l_2}
(g\dot{G}G)
\frac{j_{l_1}(k_1\rad)}{k_1}
j_{l_2}(k_2\rad)
\delta_\delta^\lin({\bf k_1})\delta_g({\bf k_2})
\nonumber\\
&& \times
Y_{l_1}^{m_1}(\hat{\veck}_1) Y_1^{m'}(\hat{\veck}_1)
Y_{l_2}^{m_2}(\hat{\veck}_2)  I_{l l_1 l_2 1}^{m^\ast m_1^\ast
m_2^\ast m'^{\ast}}(\hat{\bf n}) \, .
\label{eqn:almdsz}
\end{eqnarray}
Here and throughout, we make use of the general integral over
spherical harmonics written such that
\begin{equation}
I_{l_1 l_2 ... l_i}^{m_1 m_2 ... m_i}(\hat{\bf n}) = \int d\hat{\bf n}
Y_{l_1}^{m_1}(\hat{\bf n}) Y_{l_2}^{m_2}(\hat{\bf n})
.... Y_{l_i}^{m_i}(\hat{\bf n})
\end{equation}

We can construct the angular power spectrum by considering
$\langle a_{l_1m_1} a^*_{l_2m_2} \rangle$.
Under the  assumption that the temperature field is statistically
isotropic, the correlation is independent of $m$, and we can write the
angular power spectrum as
\begin{eqnarray}
\langle \alm{1}^{*, \dsz} \alm{2}^\dsz\rangle = \deld_{l_1 l_2} \deld_{m_1 m_2}
        C_{l_1}^\dsz\, .
\end{eqnarray}

The correlation can be written using
\begin{eqnarray}
&& \langle a^{\ast, \dsz}_{l_1m_1} a^\dsz_{l_2m_2} \rangle= \frac{(4 \pi)^6}{9}
\int d\rad_1 g \dot{G} G \int d\rad_2 g \dot{G} G  \nonumber \\
&\times& \int \frac{d^3{\bf k_1}}{(2\pi)^3}\frac{d^3{\bf k_2}}{(2\pi)^3}
\frac{d^3{\bf k_1'}}{(2\pi)^3}\frac{d^3{\bf k_2'}}{(2\pi)^3}
\nonumber \\
&&\sum_{l_1'm_1' l_1'' m_1'' m_1''' l_2'm_2' l_2'' m_2''
m_2'''} \langle \delta_\delta^\lin({\bf k_1'})\delta_g({\bf k_2'})
\delta_\delta^{\ast \lin}({\bf k_1})\delta_g^\ast({\bf k_2})  \rangle 
\nonumber \\
&\times& (-i)^{l_1'+l_1''} (i)^{l_2'+l_2''}
j_{l_2'}(k_1'\rad_2) \frac{j_{l_2''}(k_2'\rad_2)}{k_2'}
\frac{j_{l_1'}(k_1\rad_1)}{k_1}
j_{l_1''}(k_2\rad_1) \nonumber \\
&\times& 
 Y_{l_2'}^{m_2'}(\hat{\veck_1'}) Y_{1}^{m_2'''}(\hat{\veck_2'})
Y_{l_2''}^{m_2''}(\hat{\veck_1'}) 
Y_{l_1'}^{m_1' \ast}(\hat{\veck_1})
Y_{1}^{m_1''' \ast}(\hat{\veck_1}) Y_{l_1''}^{m_1'' \ast}(\hat{\veck_2}) \nonumber \\
&\times& I_{l_2 l_2' l_2'' 1}^{m_2 m_2'^{\ast} m_2''^{\ast} m_2'''^{\ast}}(\hat{\bf m}) I_{l_1 l_1' l_1'' 1}^{m_1^\ast m_1' m_1'' m_2'''}
(\hat{\bf n}) \, .
\nonumber \\
\end{eqnarray}
We can separate out the contributions such that the total is made of
correlations
following $\langle v_g v_g\rangle \langle \delta_g \delta_g \rangle$
and $\langle v_g \delta_g \rangle \langle v_g \delta_g \rangle$
depending on
whether we consider cumulants by combining $\veck_1$ with $\veck_1'$
or $\veck_2'$ respectively. After some straightforward but tedious
algebra, and noting that 
\begin{equation} 
\sum_{m_1' m_2'} 
\left(
\begin{array}{ccc}
l_1' & l_2' & l_1 \\
m_1' & m_2'  &  m_1
\end{array}
\right)
\left(
\begin{array}{ccc}
l_1' & l_2' & l_2 \\
m_1' & m_2'  &  m_2
\end{array}
\right) 
= \frac{\delta_{m_1 m_2} \delta_{l_1 l_2}}{2l_1+1}
\end{equation} 
we can write
\begin{eqnarray}
&&C_l^\dsz = \frac{2^2}{\pi^2} \sum_{l_1 l_2}
\left[\frac{(2l_1+1)(2l_2+1)}{4\pi}\right]
\left(
\begin{array}{ccc}
l & l_1 & l_2 \\
0 & 0  &  0
\end{array}
\right)^2 \nonumber \\
&\times& \int d\rad_1 g \dot{G} G
\int d\rad_2 g \dot{G} G
\int k_1^2 dk_1 \int k_2^2 dk_2 \nonumber \\
&\times& \Big( P_{\delta\delta}^\lin(k_1)
P_{gg}(k_2)j_{l_1}(k_2\rad_2) j_{l_1}(k_2\rad_1) 
\frac{j_{l_2}'(k_1\rad_1)}{k_1} \frac{j_{l_2}'(k_1\rad_2)}{k_1}
\nonumber \\
&+& P_{\delta g}(k_1) P_{\delta g}(k_2) j_{l_2}(k_2\rad_1)
\frac{j_{l_1}'(k_1\rad_1)}{k_1}j_{l_1}(k_1\rad_2)
\frac{j_{l_2}'(k_2\rad_2)}{k_2}
  \Big) \, . \nonumber \\
\end{eqnarray}

Here, the first term represents the contribution from $\langle v_g v_g
\rangle
\langle \delta_g  \delta_g \rangle$  while the second term is
the $\langle v_g \delta_g \rangle \langle v_g \delta_g \rangle$
contribution, respectively.
In simplifying the integrals involving spherical harmonics, we have
made use of the properties of Clebsh-Gordon coefficients, in
particular, those involving $l=1$.
The integral involves two distances
and two Fourier modes and is summed over the Wigner-3$j$ symbol to
obtain the power spectrum. The above equation represents 
the angular power spectrum of kinetic SZ, or OV effect, under the
all-sky corrdinates and we have not used the flat-sky assumption or
any small angular scale limit which is usually considered in the literature.

Since we are primary interested in the contribution at small
angular scales here, we can ignore the contribution to the kinetic SZ effect
involving the correlation between linear density field and baryons and
only
 consider the contribution that results from baryon-baryon and
density-density
correlations. In fact, under the halo description provided here, there
is no correlation of the baryon field within halos and the velocity
field traced by individual halos. Thus, contribution to the
baryon-velocity correlation only comes from the 2-halo term of the
density field-baryon correlation. This correlation is suppressed at
small scales and is not a significant contributor to the kinetic SZ
power spectrum (see, \cite{Hu00a}). 

Similar to the Limber approximation \cite{Lim54},
in order to simplify the calculation associated with $\langle v_g v_g
\rangle \langle \delta_g \delta_g \rangle$, we use an equation
involving completeness of spherical Bessel functions:
\begin{equation}
\int dk k^2 F(k) j_l(kr) j_l(kr')  \approx {\pi \over 2} \da^{-2}
\deld(r-r')
                                                F(k)\big|_{k={l\over
d_A}}\,,
\end{equation}
where the assumption is that $F(k)$ is a slowly-varying function.
Applying this to the integral over $k_2$ gives
\begin{eqnarray}
&&C_l^\dsz = \frac{2}{\pi} \sum_{l_1 l_2}
\left[\frac{(2l_1+1)(2l_2+1)}{4\pi}\right]
\left(
\begin{array}{ccc}
l & l_1 & l_2 \\
0 & 0  &  0
\end{array}
\right)^2 \nonumber \\
&\times& \int d\rad_1 \frac{(g \dot{G})^2}{d_A^2}
\int k_1^2 dk_1 
P_{\delta\delta}^\lin(k_1) P_{gg}\left[ \frac{l_1}{d_A}; \rad_1 \right]
\left(\frac{j_{l_2}'(k_1\rad_1)}{k_1}\right)^2\, . \nonumber \\
\label{eqn:redallsky}
\end{eqnarray}

The alternative approach, which has been the calculational method in many
of the previous papers \cite{OstVis86,DodJub95,Hu00a}, 
is to use a specific coordinate frame with the z-axis along $\vec{\bf k}$.
This allows one to simplify the SZ kinetic power spectrum to:
\begin{eqnarray}
&&C_l^\dsz = \frac{1}{8\pi^2} \int d\rad \frac{(g \dot{G}G)^2}{d_A^2} P_{\delta\delta}(k)^2
I_v\left(k=\frac{l}{d_A}\right) \, ,
\end{eqnarray}
with the mode-coupling integral given by
\begin{eqnarray}
&&I_v(k) = \nonumber \\
&&\int dk_1 \int_{-1}^{+1} d\mu \frac{(1-\mu^2)(1-2\mu
y_1)}{y_2^2} \frac{P_{\delta\delta}(k y_1)}{P_{\delta\delta}(k)}
\frac{P_{\delta\delta}(k y_2)}{P_{\delta\delta}(k)} \, . \nonumber \\
\end{eqnarray}
We refer the reader to \cite{DodJub95} 
for details on this derivation. In above, 
$\mu = \hat{\bf k} \cdot \hat{\bf k_1}$, $y_1 = k_1/k$ and
$y_2 = k_2/k = \sqrt{1-2\mu y_1+y_1^2}$. This flat-sky approximation
makes use of the Limber approximation \cite{Lim54}  to further simplify
the calculation with the replacement of $k = l/d_A$. The power spectra here
represent the baryon field power spectrum 
and the velocity field power spectrum; the former assumed to trace the
dark matter density field while the latter  is generally 
related to the linear dark matter density field through the use of linear
theory arguments. 

The correspondence between the flat-sky and all-sky
formulation can be obtained by noting that in the small scale limit
contributions to the flat-sky effect comes when $k_2 = |\veck -
\veck_1| \sim k$ such that $y_1 \ll 1$. In this limit, the flat sky
Ostriker-Vishniac effect reduces to a simple form given by Hu in \cite{Hu00a}:
\begin{equation}
C_l^\dsz = \frac{1}{3}
\int d\rad \frac{(g \dot{G} G)^2}{d_A^2} P_{gg}(k) v_\rms^2 \, .
\label{eqn:redflatsky}
\end{equation}
Here, $v_\rms^2$ is the rms of the uniform bulk velocity
from
large scales
\begin{equation}
v_\rms^2 = \int dk \frac{P_{\delta\delta}(k)}{2\pi^2} \, .
\end{equation}
The $1/3$ arises from the fact that rms in each component is $1/3$rd
of the total velocity.
 
In the same small scale limit, to be consistent with the flat sky
expression, we can reduce the all-sky expression
such that contributions come from a term that looks like
\begin{eqnarray}
&&C_l^\dsz =\int d\rad\frac{(g \dot{G})^2}{d_A^2}
P_{gg}\left[ \frac{l}{d_A}; \rad_1 \right] \frac{1}{3} v_{\rm rms}^2
\, .
\end{eqnarray}

A comparison of the reduced all-sky (Eq.~\ref{eqn:redallsky}) 
and flat-sky (Eq.~\ref{eqn:redflatsky}) formula in the
small-scale
limit suggests that the correspondence between the two arises when
\begin{equation}
\sum_{l_1 l_2}
(2l_1+1)(2l_2+1)
\left(
\begin{array}{ccc}
l & l_1 & l_2 \\
0 & 0  &  0
\end{array}
\right)^2 
\left[j_{l_2}'(k\rad)\right]^2 =\frac{1}{3} \, .
\label{eqn:simplify}
\end{equation}
Numerically, we determined this to be true as long as $l_2 \ll l$,
however, we have not been able to prove
this relation analytically. We leave this as a challenge to our readers.
Note that the limit $l_2 \ll l$ is the small-scale limit in the
flat-sky and is effectively equivalent to $y_1 \ll 1$ considered in
writing Eq.~\ref{eqn:redflatsky}.
Thus, the right hand side of the above
expression denotes the all-sky equivalent of the integral that
leads to a 1/3rd of rms of a randomly directed quantity along one
particular line of sight.

\begin{figure}[t]
\centerline{\psfig{file=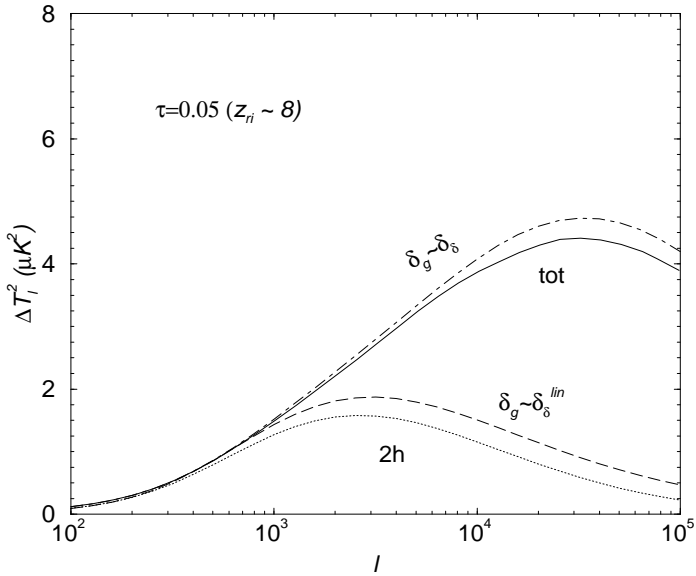,width=3.6in,angle=-90}}
\caption{The temperature fluctuation power $(\Delta T^2_l =
l(l+1)/(2\pi) C_l T_{\rm CMB}^2)$ for a variety of methods to
calculate the kinetic SZ effect. Here, we show the contribution for a
reionization redshift of $\sim$ 8 and an optical depth to reionization
of 0.05. The contributions are calculated under the assumption that
the baryon field traces the non-linear dark matter ($P_g(k) =
P_\delta(k)$ with $P_\delta(k)$ predicted by the halo model), 
the linear density field ($P_g(k) = P^\lin(k)$), and the halo model
for gas, with total and the 2halo contributions  shown separately.
For the most part, the kinetic SZ effect can be described using linear
theory, and the non-linearities only increase the temperature
fluctuation power by a factor of a few at $l \sim 10^5$.}
\label{fig:ovtemp}
\end{figure}

\begin{figure}[!h]
\centerline{\psfig{file=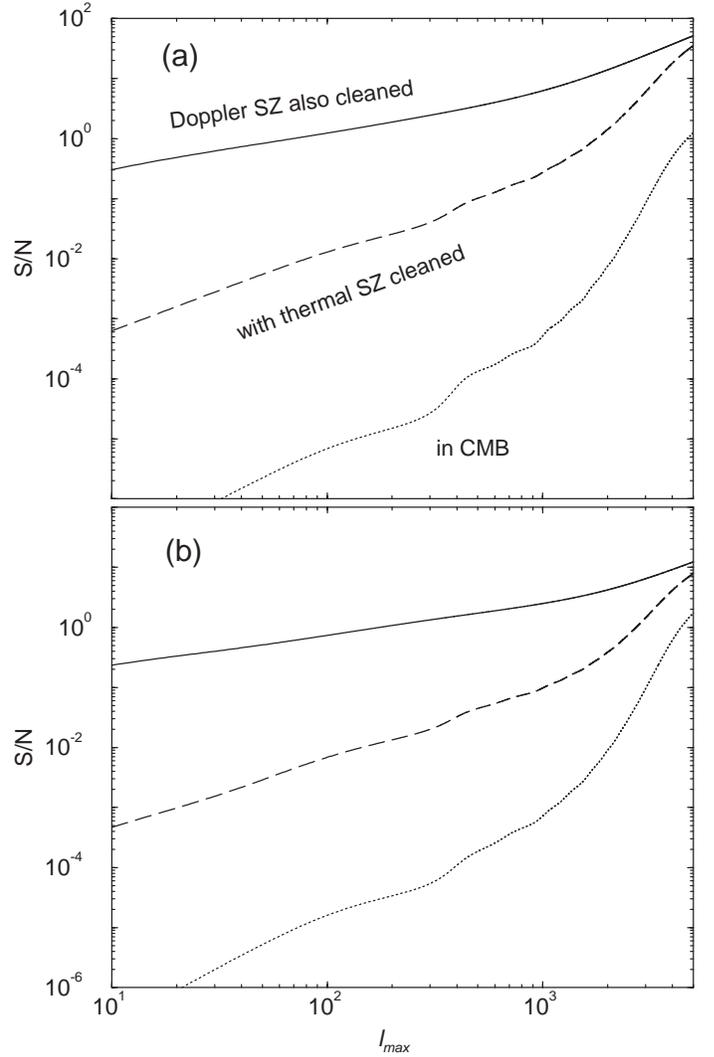,width=3.6in}}
\caption{Cumulative signal-to-noise for the detection of SZ thermal-SZ
thermal-SZ kinetic bispectrum (a) and skewness (b) with temperature anisotropy data. The
dotted line is for the detection of SZ thermal-SZ kinetic
correlation using CMB data alone,
while the dashed line is
the same when the SZ thermal effect has been separated from other CMB 
contributions and the measurement now involves two points from the SZ
map and one point from the CMB.
Finally, the solid line is the maximum signal-to-noise for achievable
with the separation of the SZ kinetic effect
from all contributors to CMB  anisotropy.}
\label{fig:bisn}
\end{figure}

\subsection{Discussion}

In Fig.~\ref{fig:szpower}, we show our prediction for the SZ kinetic
effect and a comparison with the SZ thermal contribution.
As shown, the SZ kinetic contribution is roughly an order of magnitude
smaller than the kinetic SZ contribution. 
There is also a more fundamental difference between the two:
the SZ thermal effect, due to its dependence on highest temperature
electrons is more dependent on the most massive halos in the universe,
while the SZ kinetic effect arises more clearly due to 
large scale correlations of the halos that make the
large scale structure. The difference arises from that fact that
kinetic SZ effect is mainly due to the baryons and not the temperature
weighted baryons that trace the pressure responsible for the thermal
effect. Contributions to the SZ kinetic effect comes from baryons
tracing all scales and down to small mass halos.
The difference associated with mass dependence 
between the two effects suggests that a wide-field 
SZ thermal effect map and a wide-field 
SZ kinetic effect map will be different from each other in that
massive halos, or clusters, will be clearly visible in a SZ thermal
map while the large scale structure will be more evident in
a SZ kinetic effect map.  Numerical simulations are in fact consistent with
this picture  \cite{Spretal00}.

As shown in Fig.~\ref{fig:szpower}(b), the
variations in maximum mass used in the calculation does not lead to
orders of magnitude changes in the total kinetic SZ contribution; This change
is less than the changes in the total thermal SZ
contribution as a function of maximum mass. This again is consistent with our basic
result that most contributions come from the large scale linear velocity
modulated by baryons in halos. 
Consequently, while the thermal SZ effect is dominated by shot-noise contributions, and is
heavily affected by the sample variance, the same is not true for the
kinetic SZ effect. 

In Fig.~\ref{fig:ovtemp}, we show several additional predictions for the kinetic
SZ effect, following the discussion in \cite{Hu00a}. Here, 
we have calculated the kinetic SZ power spectrum under several
assumptions, including the case when gas is assumed to trace 
the non-linear density field and the linear density field. 
We compare predictions based on such assumptions
to those calculated using the halo model. As shown, the halo model
calculation shows slightly less power than when using the non-linear dark
matter density field to describe clustering of
baryons. This difference arises from the fact that baryons do not
fully trace the dark matter in halos. Due to small differences,
one can safely use the non-linear dark matter power spectrum to describe
baryons. Using the linear theory only, however, leads to an underestimate of
power at a factor of 3 to 4 at scales corresponding to multipoles of
$l \sim 10^4$ to $10^5$ and may not provide an accurate description of
the total kinetic SZ effect.

The interesting experimental possibility here is whether one can obtain
an wide-field map of the SZ kinetic effect. Since it is now well
known that the unique spectral dependence of the thermal SZ effect can
be used to separate its contribution \cite{Cooetal00a}, at smaller
angular scales, it is likely that after the separation, SZ kinetic effect will be the
dominant signal, even after accounting for the lensed CMB
contribution. For such a separation of the SZ thermal effect to be
carried out and such that a detection of the kinetic SZ effect will be possible, 
observations, at multifrequencies, are needed to arcminute scales. 
Upcoming interferometers and similar experiments
will allow such studies to be eventually carried out. A wide-field
kinetic SZ map of the large scale structure will eventually allow an understating of
the large scale velocity field of baryons, as the  density fluctuations
can be identified through cross-correlation of such a map with a
similar thermal SZ map. We now discuss the existence of correlations
between the SZ thermal and SZ kinetic effect.

\section{SZ Thermal-SZ kinetic Correlation}

The SZ thermal and SZ kinetic effects both trace the large scale
structure baryons. One can study a correlation between these two
effect to probe  the manner in which baryons are distributed in the
large scale structure. For example, such a correlation study may allow one to answer
to what extent diffuse baryons contribute to thermal SZ when compared
to their contribution to kinetic SZ. Given that the SZ kinetic effect
is second order in fluctuations, there is no direct two-point
correlation function between the temperature anisotropies produced by
SZ thermal and kinetic effects. As discussed in \cite{CooHu00},
to the lowest order,  the correlation between kinetic SZ and thermal SZ manifests as a
nonvanishing bispectrum in temperature fluctuations and can be studied
by considering a three-point correlation function or a bispectrum, 
the Fourier space analog of the three point function, or associated statistics
such as the third moment or skewness.

\begin{figure}[!h]
\centerline{\psfig{file=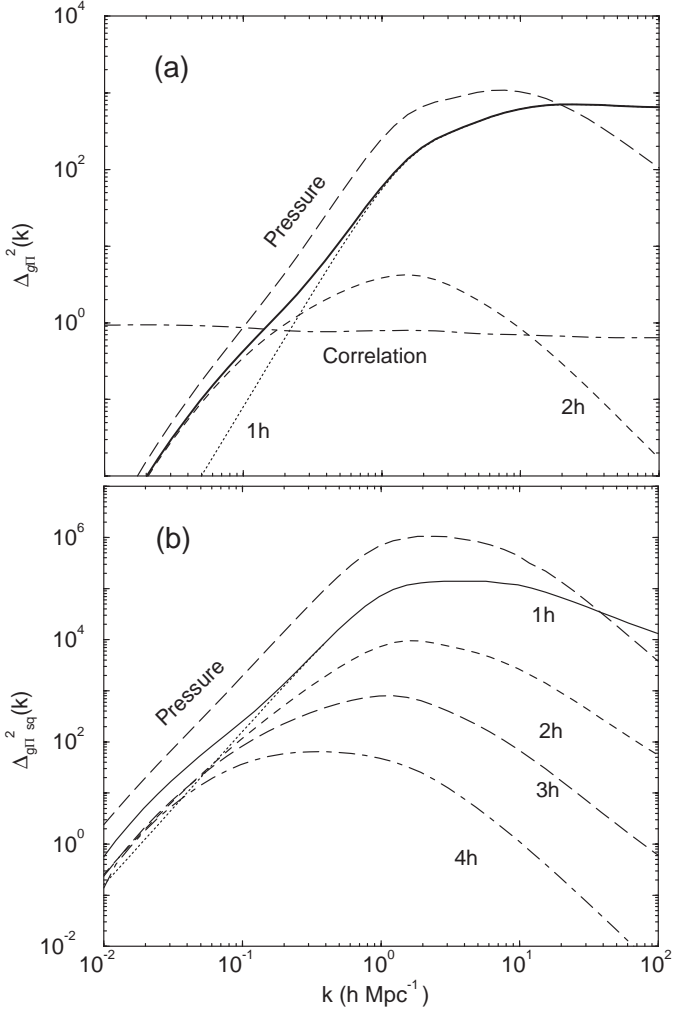,width=3.8in}}
\caption{The baryon-pressure (a) power spectrum (b) trispectrum
today ($z=0$)
broken into individual contributions under the halo description.
The line labeled 'correlation' shows the correlation coefficient of
gas-pressure correlation with respect to gas-gas and
pressure-pressure. For reference, we also show the pressure power
spectrum and the trispectrum.}
\label{fig:gaspressurepower}
\end{figure}

\subsection{SZ thermal-SZ thermal- SZ kinetic bispectrum}

We refer the reader to \cite{CooHu00} for full details on
the derivation of the SZ thermal-SZ kinetic correlation as a
bispectrum involving two measurements of SZ thermal effect and one
measurement of SZ kinetic effect (SZ thermal-SZ thermal-SZ kinetic);
note that in \cite{CooHu00}, we identify this bispectrum as SZ-SZ-OV.
As discussed there, 
for low optical depths to reionization, the detection of the SZ-SZ-OV
bispectrum is problematic due to high cosmic variance resulting from
primary anisotropies. 

In Cooray \& Hu \cite{CooHu00}, we assumed that pressure and baryons both traces dark matter
to calculate the SZ-SZ-OV bispectrum. Using our halo model, we can
now update this calculation to include the bispectrum formed between
SZ thermal-SZ thermal-SZ kinetic effects. Additionally, we can
investigate the improvements in the signal-to-noise for the bispectrum
detections when the SZ thermal effect is separated from CMB. The
maximum signal-to-noise for this bispectrum can only be achieved when the
SZ kinetic effect is also separated from CMB, though, given that the
two effects have the same frequency dependence, it is unlikely that
the kinetic SZ effect can be separated from thermal CMB anisotropies
using frequency information alone.
Using the frequency information, 
with the thermal SZ contribution separated from CMB, however, it
is likely that the SZ kinetic effect will dominate the small angular
scale signal in the temperature anisotropies. Thus, one can use small
angular scale thermal CMB temperature fluctuations to construct the SZ
thermal-SZ kinetic bispectrum. 

Following \cite{CooHu00}, we can write the SZ thermal-SZ
thermal-SZ kinetic bispectrum as
\begin{eqnarray}
\bi &=& \sqrt{\frac{\prod_{i=1}^3(2l_i +1)}{4 \pi}}
\left(
\begin{array}{ccc}
l_1 & l_2 & l_3 \\
0 & 0  &  0
\end{array}
\right) [ b^{\se-\se}_{l_1,l_2} + {\rm Perm.}] \, ,
\label{eqn:ovbidefn}
\end{eqnarray}
with
\begin{eqnarray}
&&b^{\sz-\sz}_{l_1,l_2}
=\frac{2}{\pi} \int \frac{d\rad}{d_A^2} W^\sz(\rad) g 
P_{g\Pi}\left(\frac{l_2}{d_A};\rad\right) 
\nonumber \\
&\times& 
\int d\rad_1 \int k_1 dk_1 P_{\delta\Pi}^{2h}(k_1;\rad_1) 
W^\sz(\rad_1) j_{l_1}'(k_1\rad_1) j_{l_1}(k_1\rad) \, , \nonumber \\
\label{eqn:finalintegral}
\end{eqnarray}
Here, ``Perm.'' means a sum over the
remaining 5 permutations of ($l_1$,$l_2$,$l_3$) as usual.

Here, $P_{g\Pi}$ is the baryon-pressure power spectrum while the
$P_{\delta\Pi}$ is the density-pressure power spectrum, with $\delta$
tracing the linear velocity field. Since there is no contribution to
the large scale bulk flows from the non-linear regime (ie. the 1-halo
term), we model the pressure-density cross power 
as the large scale density-pressure correlations in the
linear regime described by the 2-halo term.

The signal-to-noise for the detection of the bispectrum is
\begin{equation}
\left(\frac{{\rm S}}{{\rm N}}\right)^2 \equiv {1 \over \sigma^2(A)} =
\sum_{l_3\ge l_2 \ge l_1}
        \frac{\bi^2}{C_{l_1}^\tot C_{l_2}^\tot C_{l_3}^\tot}\,,
\label{eqn:chisq}
\end{equation}
where 
\begin{equation}
C_l^\tot = C_l^{\rm CMB}+C_l^{\rm sec}+C_l^{\rm Noise} \, .
\end{equation}

In \cite{Cooetal00a}, we showed how
multifrequency cleaning of SZ effect can be a useful tool for higher
order correlation studies and discussed how the signal-to-noise for the
detection of SZ-lensing correlation, again through a bispectrum, can
be improved by using CMB primary anisotropy separated SZ map. We
consider the same approach here, where we study the possibility for a
detection of the SZ thermal-SZ kinetic correlation by using a
frequency cleaned SZ thermal map, which will provide two measurements,
and a CMB map containing only the CMB primary, SZ
kinetic and and thermal-like secondary contributions.

In Fig.~\ref{fig:bisn}(a),
 we update signal-to-noise results for the bispectrum 
given in \cite{CooHu00}, where
we only studied the possible detection in CMB data alone and with no
consideration for separation of effects, especially the SZ thermal
effect.  The separation allows a
decrease in cosmic variance, as the noise is no longer dominated by
CMB primary anisotropies. This leads to an increase in the cumulative
signal-to-noise. With SZ thermal effect separated, we see that the
signal-to-noise increases by roughly two orders of magnitude. In
\cite{Cooetal00a}, we showed how one can obtain an order of magnitude
in signal-to-noise when a CMB separated SZ thermal map is used for a
detection of SZ-lensing correlation. Here, we obtain roughly two orders of
magnitude improvement, since the SZ thermal-SZ kinetic correlation is
present with two SZ thermal measurements, instead of one in the case
SZ thermal-lensing correlation through the SZ thermal-CMB-CMB bispectrum.

If one can separate the SZ kinetic such that a perfect SZ kinetic map,
as well as a perfect SZ thermal map, is available, then one can improve the
signal-to-noise for detection significantly such that a detection is
possible. Since the SZ kinetic effect is expected to dominate
temperature anisotropies at
small angular scales, when SZ thermal is removed, an opportunity to
detect the SZ thermal-SZ kinetic correlation will likely come from
small angular scale multifrequency experiments. One can also improve
the possibility of detecting this correlation by noting that the
configuration for the bispectrum is such that it peaks for highly
flattened triangles (see, \cite{CooHu00}). 
It is likely that progress in experimental studies will
continue to a level where such studies will eventually be possible.

\begin{figure}[!h]
\centerline{\psfig{file=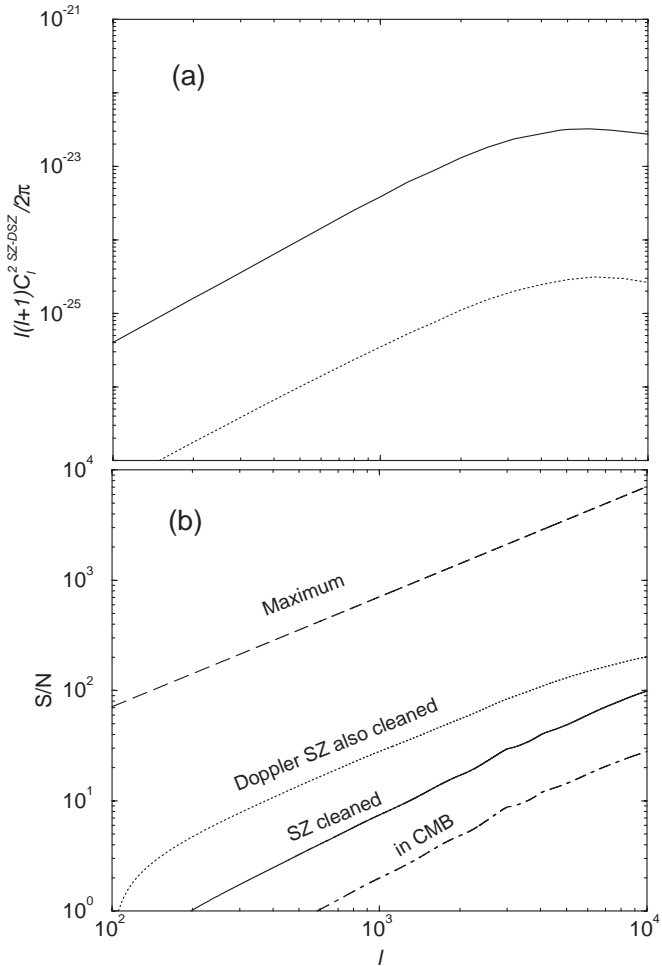,width=3.6in}}
\caption{ (a) The SZ thermal-SZ kinetic power spectrum of squared
temperatures. Here, we show the contribution to the power spectrum
when only Gaussian terms, ie. power spectra of pressure-density
correlation, are considered (dotted line) and when 
when non-Gaussianities are introduced through the
pressure-density trispectrum (solid line). In (b), we show the
cumulative signal-to-noise for the detection of the SZ thermal-SZ
kinetic squared temperature power spectrum using information in
multipoles from 2000 to 10000 and assuming no instrumental or any
other noise contributions to the covariance. The signal-to-noise is calculated
assuming the power spectrum is measured in CMB data (dot-dashed line),
with a perfect frequency cleaned SZ thermal map (solid line) and with a
perfect SZ thermal and SZ kinetic effect maps (dotted line). In a
dashed line, we show the maximum signal-to-noise achievable for the
power spectrum, with only a Gaussian contribution to the covariance.}
\label{fig:sz2ov2}
\end{figure}	

Since the bispectrum may be hard to measure directly from observational data,
we also consider a real space statistic that probes the non-Gaussian
information at the three point level and consider the third moment
(see, \cite{Cooetal00a} for details). 
In Fig.~\ref{fig:bisn}(b), we show the signal to noise for the
detection of the third moment. Here, we use a
top-hat window in multipole space out to $l_{\rm max}$ so that direct
comparison is possible with the signal-to-noise calculation involving
the bispectrum. As shown, we find that there is less
signal-to-noise in the skewness when compared to the full bispectrum. 
This results from the fact that bispectrum contains all
information at the three point level, while a measurement of the third moment
leads to a loss of information. This can also be understood by
noting that the signal-to-noise for the bispectrum and skewness is
such that in the case of the bispectrum signal-to-noise is calculated
for each mode and summed up while for the skewness signal-to-noise is
calculated after summing the signal and noise separately over all modes.

\subsection{The SZ Thermal$^2$-SZ kinetic$^2$ Power Spectrum}

In addition to the SZ thermal-SZ kinetic-SZ kinetic bispectrum, we can
introduce higher order correlations involving the SZ thermal and SZ
kinetic effect that probe the correlation between the two.
One such a possibility is the trispectrum formed by the SZ thermal and
SZ kinetic effect. 
Given that we do not have a reliable method to measure the bispectrum even,
the measurement of such a higher order correlations in experimental
data is likely to be challenging. 

Here, we focus on a statistic
that captures the correlation information coming from 
higher order, essentially from a trispectrum, but is easily measurable in
experimental data since it only involves only 
a power spectrum. Such a possibility
involves the power spectrum of squared
temperatures instead of the usual temperature itself. 
Our motivation for such a statistic came when we inspected the
published maps of the large scale SZ thermal and SZ kinetic
effects in simulations by \cite{Spretal00} and realized
that there is a significant correlation between the two
effects. Since the temperature fluctuations produced by the SZ kinetic effect 
oscillate between positive and negative values depending on the
direction of the velocity field along the line of sight, as stated earlier,
a direct two point correlation involving the temperature results in
no contribution.  A non-zero correlation between the SZ thermal and SZ
kinetic effects still manifests if the absolute
value of the temperature fluctuation due to kinetic SZ effect is considered.
Since absolute value of temperature is equivalent to squaring the temperature,
we consider the cross-correlation of SZ thermal and SZ kinetic effects
involving the power spectrum of squared temperatures here.

In order to calculate the SZ thermal$^2$-SZ kinetic$^2$ 
power spectrum, we can take the all-sky approach presented to describe
the SZ kinetic effect. The end result, however, is numerically
cumbersome since it involves 5 dimensional sum over a Wigner-$6$j
symbol (see, \cite{Coo01} for a derivation). Here, we take a
flat-sky approach and derive the thermal and kinetic SZ squared power
spectrum in the limit that the velocity field of the kinetic effect is
independent of the baryon fluctuations.

\begin{figure}[!h]
\centerline{\psfig{file=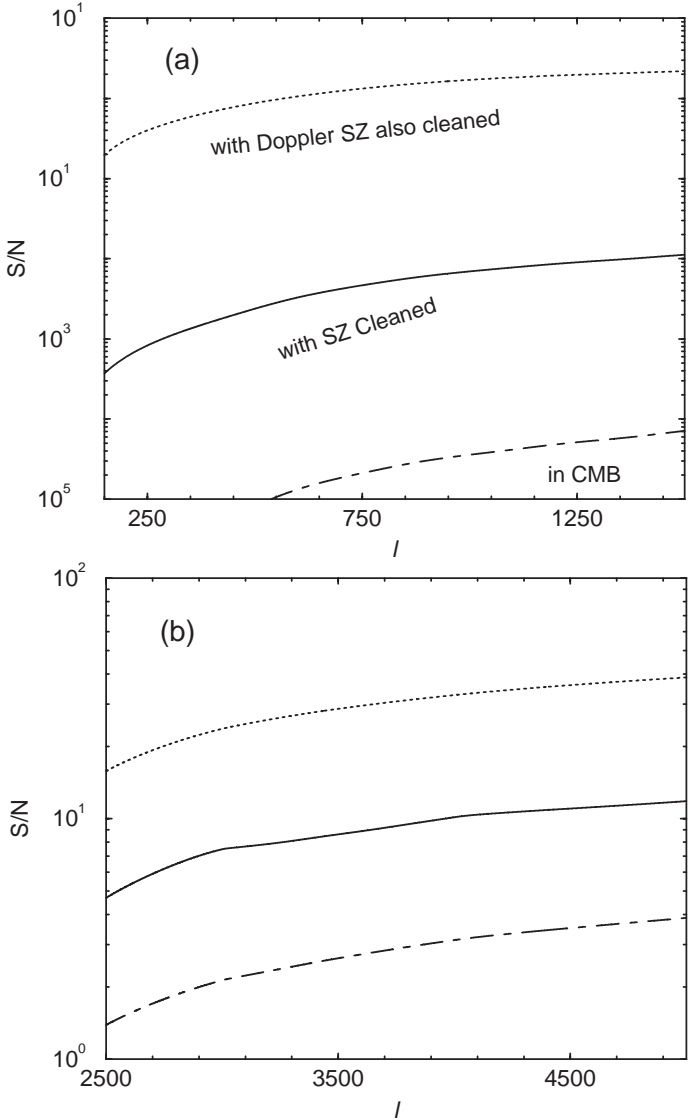,width=3.6in}}
\caption{The cumulative signal-to-noise for the detection of the thermal
SZ-kinetic SZ squared temperature power spectrum. In (a), we consider
a large angular scale experiment, consistent with Planck, and use
multipole information from $l$ of 100 to 1500. The cumulative
signal-to-noise, even with a perfectly cleaned SZ map in this
multipole range, is significantly less than 1, suggesting that a
detection is not possible. In (b), we show the cumulative
signal-to-noise for a small angular scale experiment with information
in the multipole range of 2000 to 5000. There is adequate
signal-to-noise for the detection of the squared temperature power
spectrum at such small scales even if the SZ effect is not completely
separated from the thermal CMB contribution. For comparison, in dotted
lines, we show the signal-to-noise achievable if the kinetic SZ effect
is separated from CMB, in addition to a SZ separated map.}
\label{fig:sz2ov2planck}
\end{figure}

Following our previous
definitions, we define the flat sky temperature squared power spectrum
as
\begin{equation}
\langle \cmb^{2\dsz}(\vecl) \cmb^{2\sz}(\vecl') \rangle = (2\pi)^2
\delta_D(\vecl+\vecl') C_l^{2\dsz-\sz} \, ,
\end{equation}
where the Fourier transform of the squared temperature can be
written as a convolution of the temperature transforms
\begin{equation}
\cmb^2(\vecl) = \int \frac{d^2\vecl_1}{(2\pi)^2}
\cmb(\vecl_1)\cmb(\vecl-\vecl_1) \, .
\end{equation}
Here, it should be understood that $\cmb^2(\vecl)$ refers to the
Fourier transform of the square of the temperature rather than square of the
Fourier transform of temperature, $[\cmb(\vecl)]^2$. 
To compute the square of the SZ thermal and
SZ kinetic temperature power spectrum, we take
\begin{eqnarray}
&&\langle \cmb^{2 \dsz}(\vecl) \cmb^{2 \sz}(\vecl') \rangle = (2\pi)^2
\delta_D(\vecl+\vecl') C_l \nonumber \\
&=& \int \frac{d^2\vecl_1}{(2\pi)^2}\int \frac{d^2\vecl_2}{(2\pi)^2}
\nonumber \\
&\times& \langle \cmb^\dsz(\vecl_1)\cmb^\dsz(\vecl-\vecl_1) 
\cmb^\sz(\vecl_2)\cmb^\sz(\vecl'-\vecl_2) \rangle \, .\nonumber \\
\label{eqn:flat}
\end{eqnarray}
Note that the Fourier transform of 
the temperature fluctuations in the flat sky is
\begin{equation}
\cmb(\vecl) = \int d^2\theta e^{-i\vecl \cdot \theta} T(\theta) \, .
\end{equation}

In the small scale limit, where the density field is separated from the
velocity field, we can separate the above cumulant 
to two parts 
with one involving just power spectra of pressure-baryon
cross-correlation and another with the trispectrum of pressure and
baryon fields. 
We first write down the Gaussian-like piece as
\begin{eqnarray}
&& \langle \cmb^\dsz(\vecl_1)\cmb^\dsz(\vecl-\vecl_1) 
\cmb^\sz(\vecl_2)\cmb^\sz(\vecl'-\vecl_2) \rangle^\g =  \nonumber \\
&& \int \frac{d\rad_1}{d_A^4} \int \frac{d\rad_2}{d_A^4} 
\left[g(\rad_1) W^\sz(\rad_2) \right]^2 
\frac{1}{3} v_{\rm rms}^2 \nonumber \\
&\times& \int \frac{dk_1}{(2\pi)} \int \frac{dk_2}{(2\pi)} 
e^{i(k_1 \rad_1+k_2 \rad_2)}
\left\langle \delta_g\left[\frac{\vecl_1}{d_A},k_1\right] 
\delta_\Pi\left[\frac{\vecl_2}{d_A},k_2\right] \right\rangle \nonumber \\
&\times& \int \frac{dk_3}{(2\pi)} \int \frac{dk_4}{(2\pi)} 
e^{i(k_3 \rad_1+k_4 \rad_2)} \nonumber \\
&\times&\left\langle \delta_g\left[\frac{\vecl-\vecl_1}{d_A},k_3\right] 
\delta_\Pi\left[\frac{\vecl'-\vecl_2}{d_A},k_4\right] \right\rangle \, ,\nonumber \\
\end{eqnarray}
where we have taken $\langle (\hat{\bf{\theta}} \cdot {\bf v})
(\hat{\bf{\theta}}' \cdot {\bf v'}) \rangle \sim 1/3 v_{\rm rms}^2$
with $1/3$ coming from the fact that only a third of the velocity
component contribute to the line of sight rms. 
We can now introduce power spectra in above
correlators such that 
\begin{eqnarray}
&& \langle \cmb^\dsz(\vecl_1)\cmb^\dsz(\vecl-\vecl_1) 
\cmb^\sz(\vecl_2)\cmb^\sz(\vecl'-\vecl_2) \rangle^\g = \nonumber \\
&& \int \frac{d\rad_1}{d_A^4} \int \frac{d\rad_2}{d_A^4} 
\left[g(\rad_1) W^\sz(\rad_2) \right]^2 
\frac{1}{3}v_{\rm rms}^2 \nonumber \\
&\times& \int \frac{dk_1}{(2\pi)} e^{ik_1 \left(\rad_1 - \rad_2\right)}
(2\pi)^2\delta_D\left(\frac{\vecl_1}{d_A}+\frac{\vecl_2}{d_A}\right)
P_{g\Pi}\left[\sqrt{\frac{l_1^2}{\rad_1^2}+k_1^2}\right] \nonumber \\
&\times&\int \frac{dk_3}{(2\pi)} e^{ik_3 \left(\rad_1 - \rad_2\right)}
(2\pi)^2\delta_D\left(\frac{\vecl-\vecl_1}{d_A}+\frac{\vecl'-\vecl_2}{d_A}\right)
\nonumber \\
&\times&
P_{g\Pi}\left[\sqrt{\frac{|l-l_1|^2}{\rad_1^2}+k_3^2}\right]\, .\nonumber \\
\end{eqnarray}
The integrals
over the line-of-sight wavevectors behave such that only perpendicular
Fourier modes contribute to the projected field, such that 
$l^2/d_A^2 \gg k^2$.  This is the so-called
Limber approximation \cite{Lim54}. Doing the integral over the
wavevector, then, results in a delta function in $(\rad_1-\rad_2)$ such
that only contributions come from the same redshift.
Putting the correlator back in the power spectrum equation 
(Eq.~\ref{eqn:flat}), we now get
\begin{eqnarray}
&& C_{l}^\g = \int \frac{d^2\vecl_1}{(2\pi)^2} \int \frac{d\rad}{d_A^4}
(g\dot{G})^2 W^{\rm sz}(\rad)^2 \frac{1}{3}v_\rms^2 \nonumber \\
&\times& 2 P_{g\Pi}\left(\frac{l_1}{d_A};\rad\right)
P_{g\Pi}\left(\frac{|\vecl-\vecl_1|}{d_A};\rad\right)\, ,
\end{eqnarray}
where we have introduced a factor of 2 account for the additional permutation
involved in the baryon density-pressure correlation.

Similarly,  the non-Gaussian piece follows as
\begin{eqnarray}
&& C_{l}^\ng = \int \frac{d^2\vecl_1}{(2\pi)^2} \int
\frac{d^2\vecl_2}{(2\pi)^2} \int \frac{d\rad}{d_A^6}
(g\dot{G})^2 W^{\rm sz}(\rad)^2 \frac{1}{3}v_\rms^2 \nonumber \\
&\times& T_{g \Pi g \Pi}\left[\left(\frac{\vecl_1}{d_A}\right),
\left(\frac{\vecl-\vecl_1}{d_A}\right),
\left(\frac{\vecl_2}{d_A}\right),\left(\frac{-\vecl-\vecl_2}{d_A}\right);\rad\right]\,.\nonumber \\
\end{eqnarray}
Since we only use the single halo term to calculate $T_{g \Pi g\Pi}$,
the arguments are simply scalars and does not dependent on the
orientation of the quadrilateral. In general, however, the trispectrum
depends on the length of the four sides plus the orientation of at
least one of the diagonals.

\subsubsection{Signal-to-Noise}

In order to calculate the possibility for a detection of the thermal
SZ$^2$-kinetic SZ$^2$ power spectrum, we
need the covariance of the estimator involved with the measurement of
the squared power spectrum:
\begin{equation}
\hat{C}_l^{2\dsz-\sz} = \frac{A_f}{(2\pi)^2} \int
\frac{d^2\vecl}{A_\shell}\cmb^{2\dsz}(\vecl)
\cmb^{2\sz}(-\vecl) \, .
\end{equation}
Here, $A_\shell = \int d^2\vecl$  is the area in the 2D shell in
Fourier space over which the integral is done and $A_f$ is the total area of
the survey in Fourier space and can be written as $A_f =
(2\pi)^2/\Omega$ with a total survey area on the sky of $\Omega$.
Following \cite{Zal00}, we can write down the covariance of our
estimator as
\begin{eqnarray}
&&\Cov\left[\left(\hat{C}_l^{2\dsz-\sz}\right)^2\right] \nonumber \\
&=& \frac{A_f}{A_\shell}
\left[\left(C_l^{2\dsz-\sz}\right)^2 + C_l^{2\dsz-\dsz}
C_l^{2\sz-\sz}\right] \, , \nonumber \\
\label{eqn:sz2ov2cov}
\end{eqnarray}
where $C_l^{2\dsz-\dsz}$ is the squared power spectrum of kinetic SZ
and thermal SZ while $C_l^{2\dsz-\dsz}$ and $C_l^{2\sz-\sz}$ are the
squared power spectra of kinetic SZ and thermal SZ respectively.
Here, we assume that squared fields are Gaussian.
To calculate $C_l^{2\dsz-\dsz}$ and $C_l^{2\sz-\sz}$ we make several
assumptions: We assume that the temperature squared power spectrum
will be measured using two maps involving frequency separated SZ
contribution (which will have $\cmb^\sz+\cmb^\noise$) and a map with
kinetic SZ contribution with the CMB primary component, such that the 
will be composed of $\cmb^{\rm primary}+\cmb^\dsz+\cmb^{\noise '}$.
Following such a separation, we can write the $C_l^{2\sz-\sz}$ as
\begin{eqnarray}
&&\langle \cmb^{2\sz}(\vecl) \cmb^{2\sz}(\vecl') \rangle = (2\pi)^2
\delta_D(\vecl+\vecl') C_l^{2\sz-\sz} \nonumber \\
&=& \int \frac{d\vecl_1}{(2\pi)^2}\int \frac{d\vecl_2}{(2\pi)^2}
\langle \cmb^\sz(\vecl_1)\cmb^\sz(\vecl-\vecl_1) 
\cmb^\sz(\vecl_2)\cmb^\sz(\vecl'-\vecl_2) \rangle\nonumber \\
&=& \int \frac{d\vecl_1}{(2\pi)^2} \left [ 2 C_{l_1}^\sz
C_{|\vecl-\vecl_1|}^\sz + \int \frac{d\vecl_2}{(2\pi)^2}
T^\sz(\vecl_1,\vecl-\vecl_1,\vecl_2,-\vecl-\vecl_2)\right] \, , \nonumber \\
\label{eqn:conv}
\end{eqnarray}
and like wise for $C_l^{2\dsz-\dsz}$.
Here, contributions come from a Gaussian part involving 
SZ power spectra and a non-Gaussian part through the SZ
trispectrum. Since the primary component fluctuations dominate the
kinetic  SZ temperature, and that there is no measurable trispectrum for this
component under current adiabatic CDM predictions we ignore any
non-Gaussian contribution to $C_l^{2\dsz-\dsz}$ and write it as the
one with the Gaussian part only. This assumption is also safe at
small angular scales out to $l \sim 10^4$, under the halo model,
when $\cmb^\dsz > \cmb^{\rm primary}$ since the kinetic SZ effect can
be described for the most part using large scale correlations instead
of the single halo term.

In the Eq.~\ref{eqn:sz2ov2cov}, the ratio of $A_\shell/A_f$ is the
total number of modes that measures the squared power
spectrum independently and can be approximated such that $A_\shell/A_f =
f_\sky(2l+1)$. To calculate the signal-to-noise involved in the
detection of the squared temperature power spectrum, we consider an
optimized estimator with a weighing factor $W_l$ such that 
\begin{equation}
\hat{Y} = \sum_l W_l \hat{C}_l^{2\dsz-\sz} \, ,
\end{equation}
and write the signal-to-noise as 
\begin{equation}
\sn =
\left[\frac{\langle\hat{Y}\rangle^2}{\Cov(\hat{Y}^2)}\right]^{1/2} \, .
\end{equation}
The weight $W_l$ that maximizes the signal-to-noise is $W_l =
C_l^{2\dsz-\sz}/\Cov[(C_l^{2\dsz-\sz})^2]$ \cite{Zal00} and we
can write the required signal-to-noise as 
\begin{eqnarray}
&&\sn = \nonumber \\
&&\left[ f_{\rm sky} \sum_l (2l+1)
\frac{\left(C_l^{2\dsz-\sz}\right)^2}{
\left(C_l^{2\dsz-\sz}\right)^2 + C_l^{2\dsz-\dsz}
C_l^{2\sz-\sz}}\right]^{1/2} \, .\nonumber \\
\end{eqnarray}

\subsection{Discussion}

In Fig.~\ref{fig:sz2ov2}(a), we show  the power spectrum of squared
temperatures for the SZ thermal and SZ kinetic effects using the halo
term. Here, we have separated the Gaussian and non-Gaussian
contribution  to the squared power spectrum. As shown, the
non-Gaussian contribution to the power spectrum is significantly
higher than the Gaussian contributions. 

The Gaussian contribution to the squared power spectrum traces the
pressure-baryon density field power spectrum, which is shown in
Fig.~\ref{fig:gaspressurepower}(a) using the halo model. In the same
figure, for comparison, we also show the pressure-pressure power
spectrum and the correlation coefficient for pressure-baryon with
respect to the pressure-pressure and baryon-baryon power spectra. The
correlation behaves such that pressure and baryons trace each other
at very large scales while the correlation is decreased at small
scales due to the turnover in the pressure power spectrum. This is
equivalent to the statement that there is no low mass halo
contribution to the pressure power spectrum; these halos continue to
contribute to the baryon density field power spectrum. 

The non-Gaussian contribution to the thermal SZ-kinetic SZ squared
temperature power spectrum traces the trispectrum formed by pressure
and density field. We show this in Fig.~\ref{fig:gaspressurepower}(b)
following the halo model. For comparison, we also show the trispectrum
formed by pressure alone in the same figure. The pressure-baryon
trispectrum  is such that at large scales, corresponding to linear
scales, significant contributions come from the correlations between
halos instead of the single halo term. If there are significant
contributions coming to the squared temperature power spectrum from
such linear scales, the Gaussian part of the power spectrum should
dominate. Since all contributions to the
squared temperature power spectrum comes from small angular scales
corresponding to non-linear scales in the pressure-baryon
trispectrum, we only use the single halo contribution in calculating
the non-Gaussian part of  the squared temperature power spectrum.
In both Gaussian and non-Gaussian parts of the power spectrum,
the velocity field of the halos are taken to be the large scale bulk flows through
the linear theory.

In order to assess the maximum possibility for a measurement of the
temperature squared power spectrum involving kinetic SZ and thermal
SZ, here, we ignore the detector and beam noise contributions to the
covariance. Also, we assume full-sky experiments with $f_\sky=1$.

As  written in Eq.~\ref{eqn:conv}, the contribution to the
covariance comes as a convolution in Fourier space. Thus, even at
small angular scales corresponding to high multipoles,
noise contributions can come from large angular scales or low
wavelength modes. Such modes do not have any signal and by only
contributing to the variance, they can reduce the
effective signal-to-noise in the measurement. Since the squared power
spectrum effectively peaks at multipoles of $\sim$ $10^4$, we can
essentially ignore any contribution to the signal, as well as noise,
from multipoles less than few thousand. These are the same multipoles
in which the CMB primary anisotropies dominate, thereby, increasing
the effective noise for the measurement. In order to remove the low
multipoles,  we introduce a filtering scheme to the spherical or
Fourier transform of the temperature anisotropy measurements and
suppress the low multipole data such that the filter essentially acts
as a high pass filter above some $l > l_{\rm min}$.

In Fig.~\ref{fig:sz2ov2}(b), we show the cumulative signal to noise
for the measurement of the thermal SZ-kinetic SZ squared temperature
power spectrum. Here, we have assumed an experiment, with no
instrumental noise, such that information is only used in multipoles
of 2000 to 10000. We find significant signal-to-noise for the
detection of the squared temperature power spectrum, especially when
using a frequency cleaned SZ map with a CMB map, which has no SZ
contribution. One can in fact use the CMB map itself especially if
multifrequency information is not available for SZ separation. Since an
experiment with multipolar information out to $l \sim 10^4$ will not
readily be available, we consider two separated realistic cases,
involving a large angular scale experiment, similar to Planck, and a
small angular scale experiment similar to the ones proposed for the
study of SZ effect. We summarize our results in
Fig.~\ref{fig:sz2ov2planck}. As shown in (a), an experiment only
sensitive to multipolar information ranging from 100 to 1500 does not
have any signal-to-noise for a detection of the squared power
spectrum. With a perfect SZ separated map in this multipolar range,
the cumulative signal-to-noise for the squared power spectrum is in the
order of $\sim$ 0.01. Thus, it is unlikely that Planck data will be useful for
this study. Since we have not included any instrumental noise in
calculating signal-to-noise, the realistic signal-to-noise for Planck
would be even lower. 

Going to smaller angular scales, we find that the
signal increases significantly such that an experiment only sensitive
to the range of $l \sim 2000$ to 5000 has adequate signal-to-noise for
a detection of the squared power spectrum. A multifrequency experiment
in the arcminute scales can use its frequency  cleaned SZ map to
cross-correlate with the CMB map and obtain the squared power spectrum
with a cumulative signal-to-noise of order few tens. A comparison to
Fig.~\ref{fig:sz2ov2}(b), suggests that going to lower scales beyond
5000 increases the signal-to-noise, and this is due to the fact that
SZ thermal-SZ kinetic squared power spectrum peaks at multipoles of
$\sim$ 7000 to 8000, suggesting that for an optimal detection of the
squared power spectrum, one should also include observations out to
such high multipoles. 

\section{Summary \& Conclusions}
\label{sec:conclusion}

We have discussed non-Gaussian effects associated with local large-scale
structure contributions to the Cosmic Microwave Background  
temperature fluctuations during the reionized epoch involving the 
thermal Sunyaev-Zel'dovich effect. At low redshift, distribution
functions associated with the large scale structure properties are
non-Gaussian, primarily due to the non-linear  gravitational
evolution. The non-Gaussianities associated with the SZ effect is due to
the non-Gaussian distribution of the large scale pressure
fluctuations. Since we do not have reliable analytical methods to
calculate the higher order clustering of pressure, or as a matter of
fact, any other property associated with large scale structure
including the dark matter distribution, we have utilized the so-called
halo approach to clustering. The basic description of this technique
is that correlation functions, and their Fourier analogies, 
can be described through clustering of the property within and between
halos. The halos themselves are assumed to be clustered with respect
to the linear density field and with a mass dependent bias. We use
linear theory and its perturbations out to second order to describe
the clustering of halos. We use the halo method to describe clustering
of pressure assuming that gas is in hydrostatic equilibrium with the
dark matter density profile in halos. We use the virial equation to
describe the electron temperatures.

It is well know that the frequency dependence of the SZ effect allows
a separation of its contribution from other temperature fluctuations
and  foreground anisotropies in multifrequency CMB experiments. Using the
pressure trispectrum, the Fourier analogue of the four-point
correlation function, calculated under the halo model,
we discuss the full covariance of the SZ thermal power spectrum. The
covariance, with the inclusion of non-Gaussianities,
 allows us to properly establish the errors associated with SZ power
spectrum measurements. Thus, we can use 
this full covariance to study the astrophysical uses of
the SZ effect. Through a Fisher matrix formalism, we discuss how well
a measurement of the SZ power spectrum can be used to conclude details
on gas and temperature evolution and show that there are significant
degeneracies between gas and temperature evolution parameters. The SZ
power spectrum from Planck mission can be used to understand the
amount of baryons present in halos that contribute to the SZ effect
and to establish any nongravitational heating of electrons, such as
due to preheating with an error of 0.7 keV.

With the SZ thermal effect separated in temperature fluctuations using
its frequency information, the kinetic SZ effect, also known as the
Ostriker-Vishniac, is expected to dominate the thermal
contribution at small angular scales corresponding to multipoles of
few thousand. This effect arises from the baryon modulation of the
first order Doppler effect. The presence of the SZ kinetic effect
can be determined through a cross-correlation between the SZ thermal
and a CMB map at small scales.
Since the SZ kinetic effect is second order, however, contributions to
such a cross-correlation arise to the lower order in the form of a
three-point correlation function, or a
bispectrum in Fourier space. We suggest an additional statistic
that can be used to study the correlation between pressure traced by
the SZ thermal effect and the baryons traced by the SZ kinetic effect
involving  the temperature anisotropy
power spectrum of squared temperatures instead of the usual
temperature itself. This power spectrum probes the trispectrum formed
by the pressure-baryon cross-correlation. Through a signal-to-noise
calculation, we show that future small angular scale multi-frequency
CMB experiments, sensitive to multipoles of a few thousand,
will be able to measure the cross-correlation of
SZ thermal and SZ kinetic effect through a temperature squared
power spectrum.

\acknowledgments
I am grateful to my advisor, Wayne Hu, for
suggesting problems and calculations presented here and
in all our papers cowritten during the last two years. 
I thank other thesis committee members, John Carlstrom, Scott
Dodelson and Don York for their guidance and helpful suggestions.
Scott Dodelson is also thanked for a careful reading of this manuscript.
During the four years at Chicago, I was supported by individual grants to John
Carlstrom and Don York, a McCormick Fellowship and a Grant-In-Aid of
Research from Sigma Xi, the national science honor society.

\end{document}